\documentclass[10pt,letterpaper]{article}
\usepackage{opex3}

\usepackage{amssymb}
\usepackage{amsmath}
\usepackage{graphicx}

\begin{document}
\title{Interaction of highly focused vector beams with a metal knife-edge}
\author{P. Marchenko$^{1,2}$, S. Orlov$^{1, 2}$, C. Huber$^{1,2}$, P. Banzer$^{1,2}$, S. Quabis$^{2}$,\\ U. Peschel$^{2}$, G. Leuchs$^{1,2}$} 

\address{$^1$Max Planck Institute for the Science of Light, G\"{u}nther-Scharowsky-Str. 1, D-91058 Erlangen, Germany}
\address{$^2$ Institute of Optics, Information and Photonics, University  Erlangen-Nuremberg, Staudtstr. 7B2, D-91058 Erlangen, Germany}
\email{Sergejus.Orlovas@mpl.mpg.de}
\begin{abstract}
We investigate the interaction of highly focused linearly polarized optical beams with a metal knife-edge both theoretically and experimentally. A high numerical aperture objective focusses beams of various wavelengths onto samples of different sub-wavelength thicknesses made of several opaque and pure materials. The standard evaluation of the experimental data shows material and sample dependent spatial shifts of the reconstructed intensity distribution, where the orientation of the electric field with respect to the edge plays an important role. A deeper understanding of the interaction between the knife-edge and the incoming highly focused beam is gained in our theoretical model by considering eigenmodes of the metal-insulator-metal structure. We achieve good qualitative agreement of our numerical simulations with the experimental findings.
\end{abstract}

\section{Introduction}

The growing interest in highly focused optical vector beams requires a proper treatment of the polarization state of the beam, which strongly influences the size of the focal spot \cite{SQuab00}. In particular, the role of azimuthal and radial polarization has been investigated both theoretically \cite{KSYoug00} and experimentally \cite{RDorn03a,RDorn03b} and optimization strategies were proposed thereafter \cite{GLeuc06}. Furthermore, when focusing with high numerical aperture (NA) objectives, the symmetry of the focal spot is broken \cite{RDorn03a} for a linearly polarized beam and strong longitudinal components appear \cite{RDorn03b}. Therefore, a precise characterization of tightly focused laser beams is not just a challenge but is essential for further applications. In the literature many methods for beam characterization are described as e.g. the knife-edge method \cite{JAArn71,AHFire77}, a point scan method \cite{MBSch81} or a slit method \cite{RLMcC84,OMat91} etc. The data provided by the knife-edge method can be evaluated after the experiment in  various ways as by employing an inversion algorithm involving a linear least-square method to measure the beam's diameter \cite{JMKho83}, by performing a numerical differentiation of data in order to directly obtain the beam profile \cite{GBro85} or  by using a direct fit with an error function \cite{HRBil85}. Also other numerical approaches \cite{MACdeA09} are used to  characterize the beam. The knife-edge method is also applied to measure mechanical displacements in the nanometer range \cite{DKar06}. The application of an optical beam profiler without any moving parts using liquid-crystal displays is another example of the application of the knife-edge method \cite{MGen07} with a micro knife-edge scanner fabricated in a silicon-on-insulator substrate \cite{YChi07}. Beside a single knife-edge, also a periodical array of slits was studied recently, demonstrating the transmission anomalies of TM-polarized light \cite{YXie06}.

The background of the conventional knife-edge method is the scalar diffraction theory, so the standard knife-edge method's evaluation scheme is polarization independent. Thus, theoretically one dimensional beam scans by a knife-edge can be used in a variety of algorithms to retrieve two dimensional beam profiles with a Radon backward transform. However, when a highly focused two dimensional beam is profiled with a knife-edge, it is natural to ask ourselves, how well the conventional knife-edge method performs for vector-beams. The role of polarization in such beam measurements was studied theoretically for the knife-edge made from an ideal conductor \cite{OMat91}. The first precise measurements of the highly focused beams using the knife-edge method were performed by R. Dorn et al \cite{RDorn03a}. However, this work already shows a very first experimental indication that without careful optimization of material and knife-edge parameters the knife-edge method can be polarization sensitive and the conventional evaluation may fail. Thus, motivated by this work we performed a systematical study on how careful various parameters of knife-edges for beam reconstruction have to be selected.

The aim of our paper is to investigate in detail the interaction of highly focused linearly polarized beams with a knife-edge, made from a variety of pure materials. We investigate the beam profiling situations for two polarizations when the electric field is either parallel (p-polarization) or normal (s-polarization) to the edge. We extend previous theoretical studies on diffraction through a finite slit \cite{OMat83} and incorporate plasmonic modes \cite{SAMai05} into the model to simulate our experimental findings. Thus, our theory is based on the modal analysis in a metal-insulator-metal waveguide, which was studied in detail elsewhere \cite{BStur07,SEKoc09}. As a result, we obtain a good agreement with our experiments.

The structure of our paper is as follows. We start with the description of our experimental setup and explain the principle of our measurement. After that, we proceed with the discussion of our experimental findings. In the third chapter we discuss the physical mechanisms in the realizable knife-edge method and their effects on the outcome of measurements. Finally, we develop a theoretical model and present results of numerical simulations, which are compared with experimental findings. 

\section{The setup, principle of measurement and experimental results}
\subsection{Setup}
Most of the experiments were performed at wavelengths from $500$ up to $700$ nm using a tunable femtosecond laser system from TOPTICA. Additionally, we performed measurements at a wavelength of $780$ nm using a laser diode-based cw-system. The collimated linearly polarized Gaussian (TEM$_{00}$) laser beam was focused onto the sample using a microscope objective with NA of $0.9$. The full width at the half maximum (FWHM) of the intensity of the incoming beam was $3.1$ mm filling 86\% of the entrance pupil of the microscope objective. The sample was mounted onto a piezo stage to control its $3$D-position with nanometer accuracy.

\begin{figure}[t!]
\centering
\includegraphics[scale=0.5]{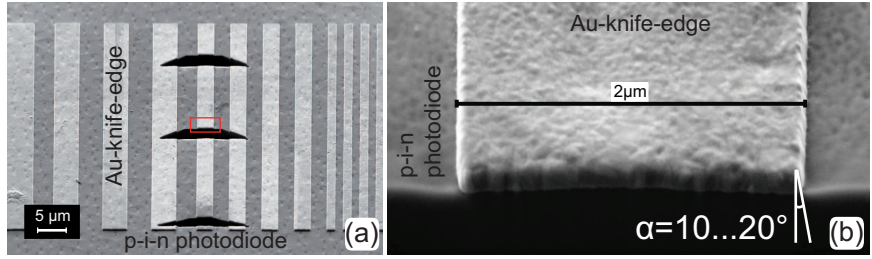} 
\caption{Electron-micrographs of one of the gold samples investigated in the experiments. The knife-edge width and film thickness were determined by performing cuts with a focused ion beam (FIB) machine. The width of the investigated knife-edge is $d=2.0$ $\mu$m, the slit width is $l=2.0$ $\mu$m.}
\label{fig:sample}
\end{figure}

As sensors we used in-house fabricated p-i-n photodiodes. Opaque and pure material films were thermally deposited on a photodiode. Therefore periodic stripe-like structures were patterned using electron-beam lithography, where the slit and the stripe width is both equal (see Fig. \ref{fig:sample}). The thickness $h$ was approx. $130$ nm. For the Au samples we additionally patterned structures with thicknesses of approx. $h=70, 100, 190$ nm. Forming structures directly on the detector surface allows the detection of a large solid angle in transmission. The materials of the opaque films were gold (Au), titanium (Ti) and nickel (Ni). To obtain the geometric parameters of the structures after finishing all measurements we cut each structure (knife-edge) by using the focused-ion-beam (FIB) technique at three positions and average values of its width at the bottom ($d=2.005$ $\mu$m) and slope angle ($\alpha=15^\circ$) were estimated (see Fig. \ref{fig:sample}). The film thickness was measured with high accuracy by means of an atomic force microscope (AFM).

\begin{figure}[t!]
\centering
\includegraphics[scale=0.33]{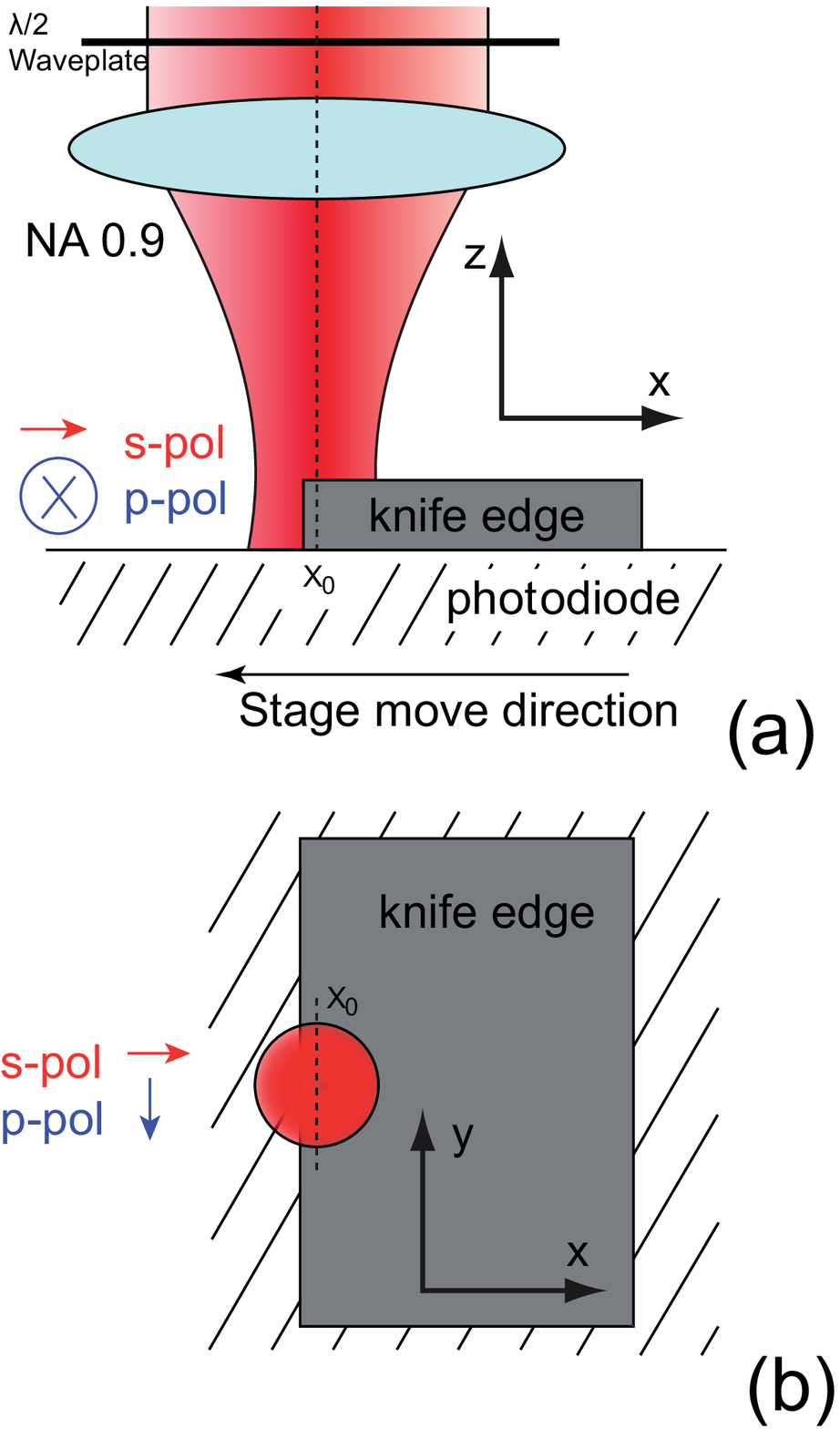}
\includegraphics[scale=0.33]{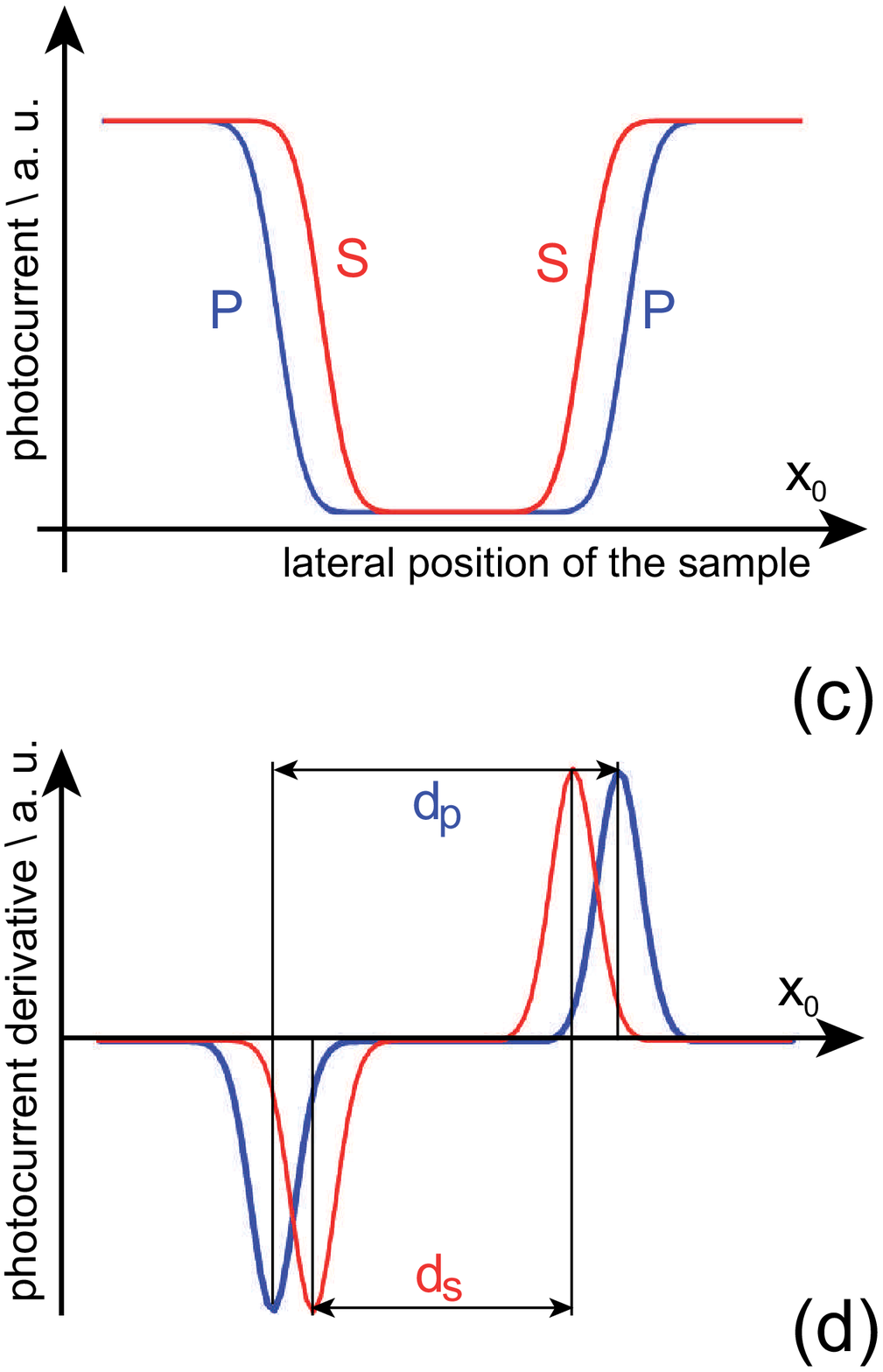} 
\caption{Schematic depiction of the knife-edge method for a two-dimensional beam (a,b). Typical beam profiling data (c) and their derivatives (d).  The state of polarization always refers to the orientation of the electric field.}
\label{fig:exp_setup}
\end{figure}

\begin{figure}[t!]
\centering
\includegraphics[scale=0.20]{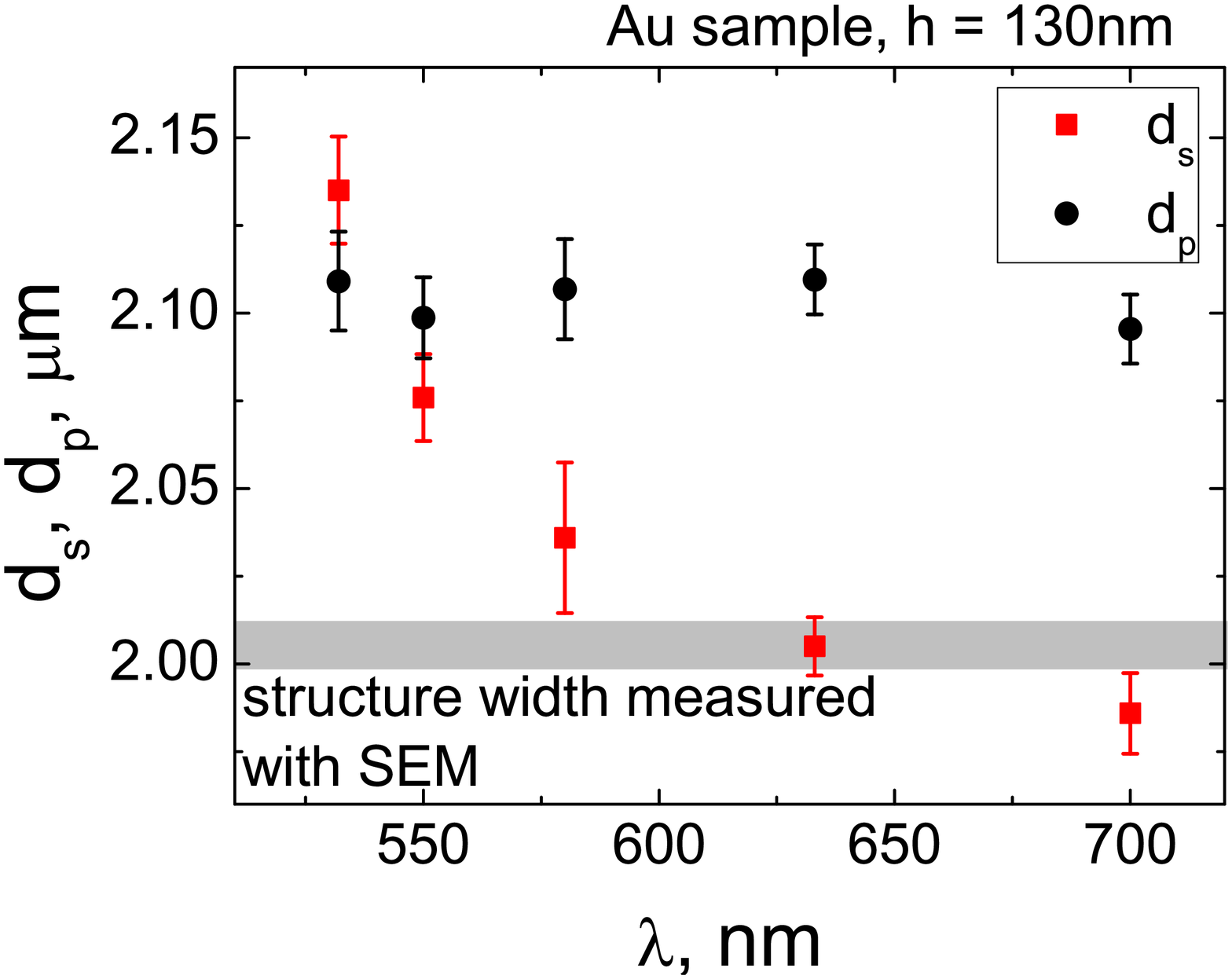} \text{(a)} 
\includegraphics[scale=0.20]{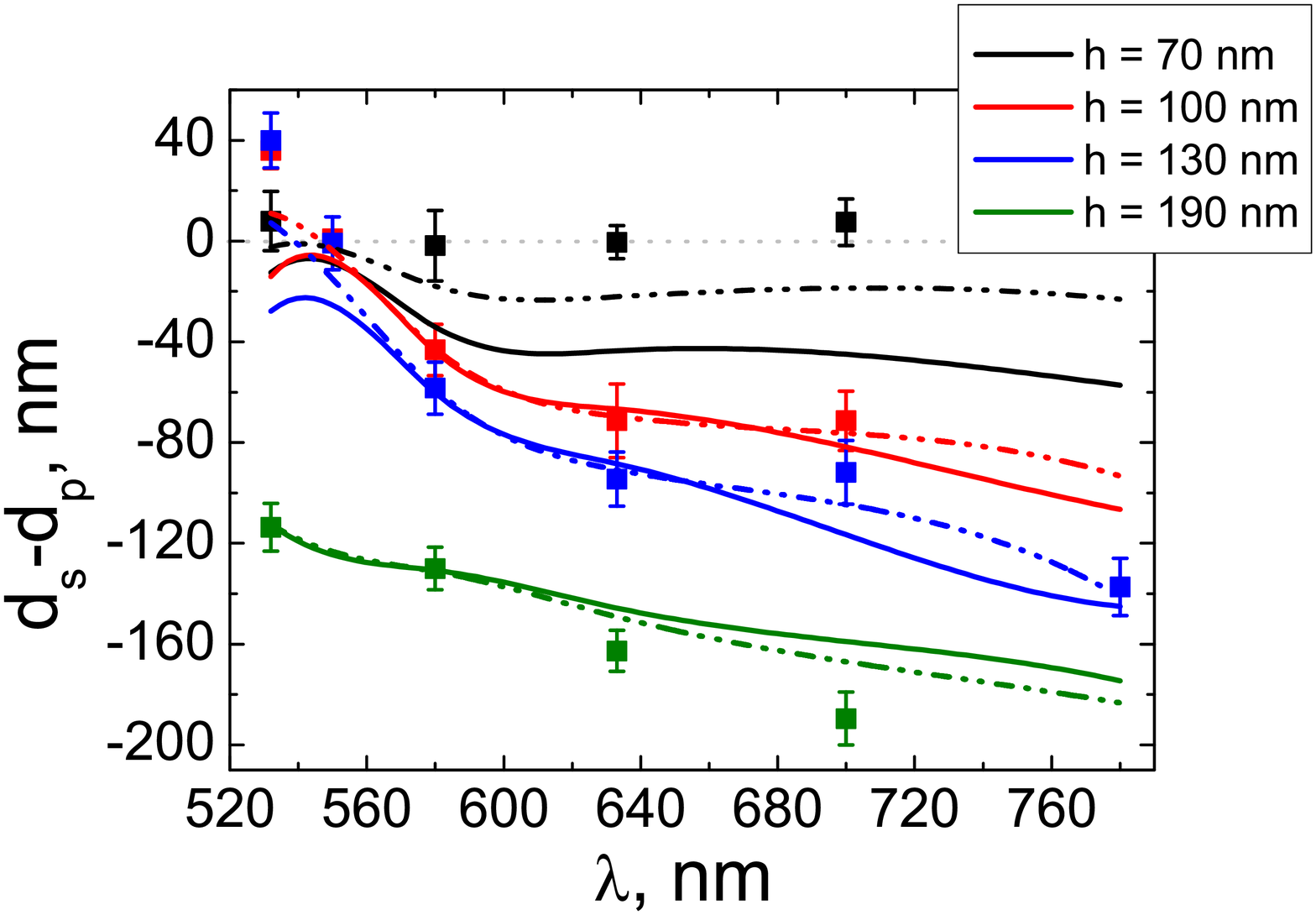}\text{(b)}
\includegraphics[scale=0.20]{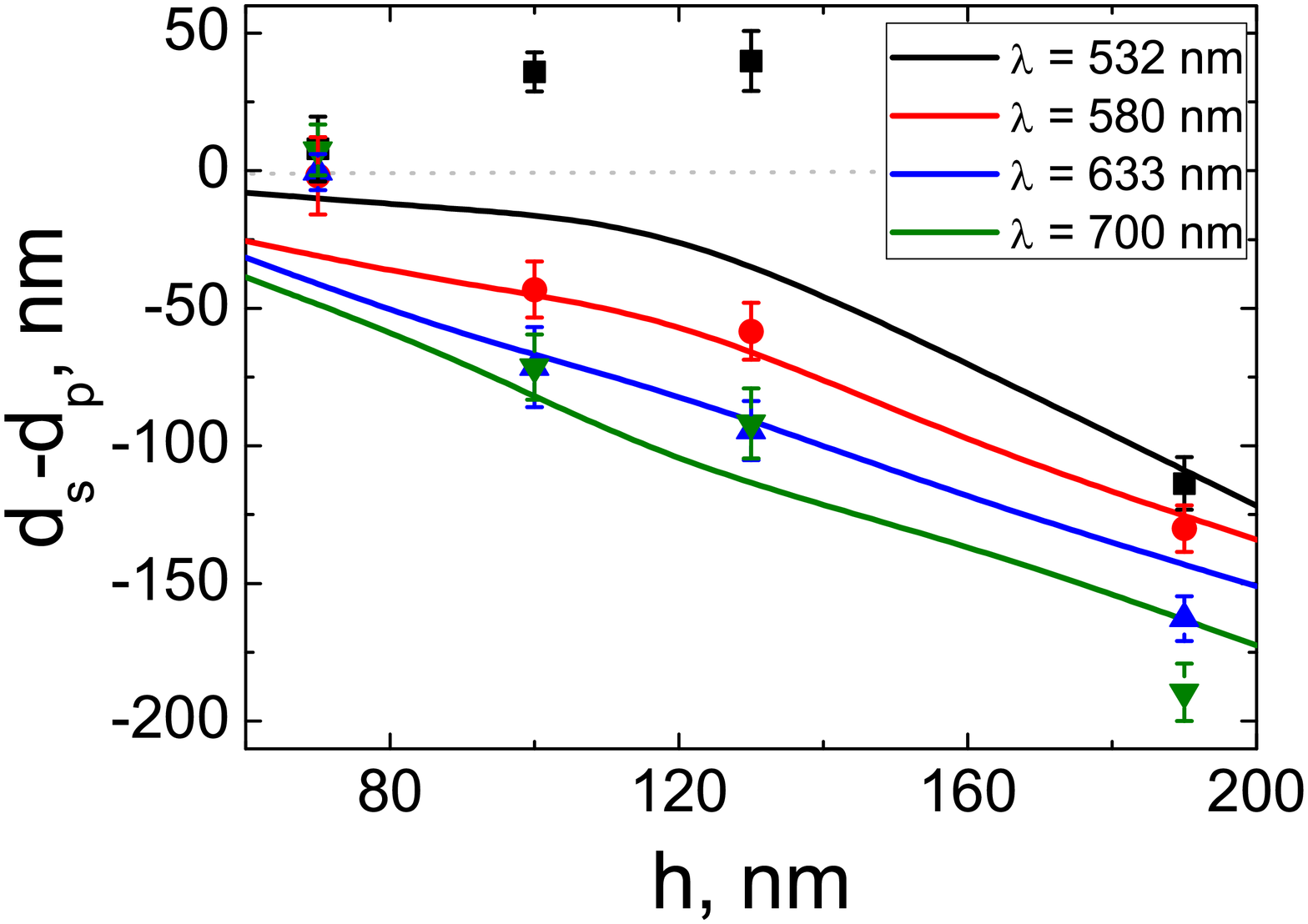}\text{(c)}
\includegraphics[scale=0.20]{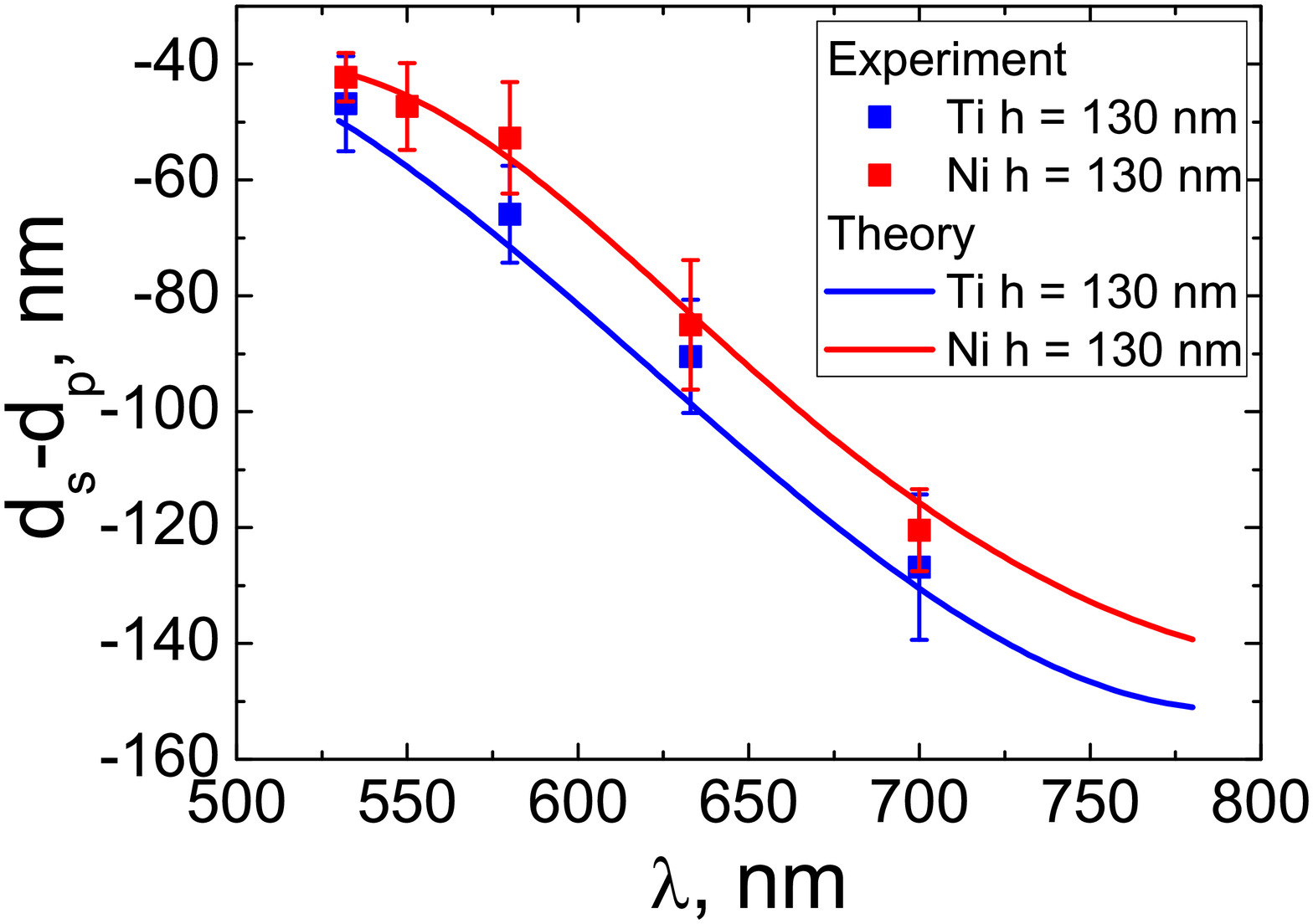}\text{(d)}
\caption{Distance between the peaks $d_s$, $d_p$ versus wavelength $\lambda$ derived from the experimental data for the s- and p-polarizations. The actual knife-edge width measured by SEM is shown by the gray bar (a). Difference of the peak positions of the photocurrent's derivative $d_s-d_p$ versus wavelength $\lambda$ for various Au samples (b). Difference of the peak positions $d_s-d_p$ of the photocurrent's derivative versus sample height $h$ for Au samples at various wavelengths (c). Difference of the peak positions $d_s-d_p$ of the photocurrent's derivative versus wavelength $\lambda$ for Ti and Ni samples (d). The colored lines represent results of numerical simulations. The dash-dotted lines in (b) represent a situation with noise artificially added to the photocurrent and smoothed with the same filter as in the experiment. The colored points represent experimental results. }
\label{fig:exp_res}
\end{figure}

\begin{figure}[t!]
\centering
\includegraphics[scale=0.2]{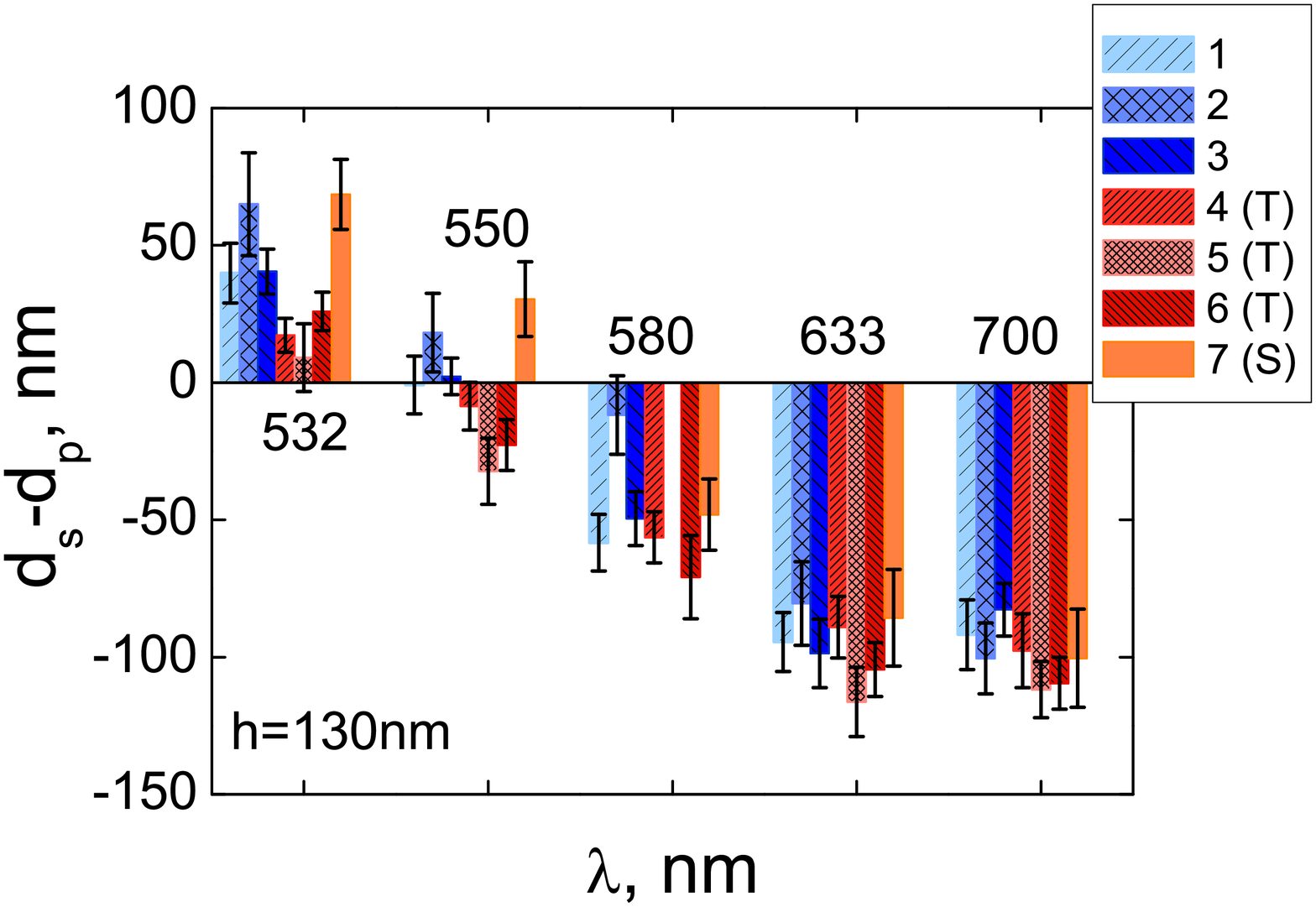} \text{(a)}
\includegraphics[scale=0.2]{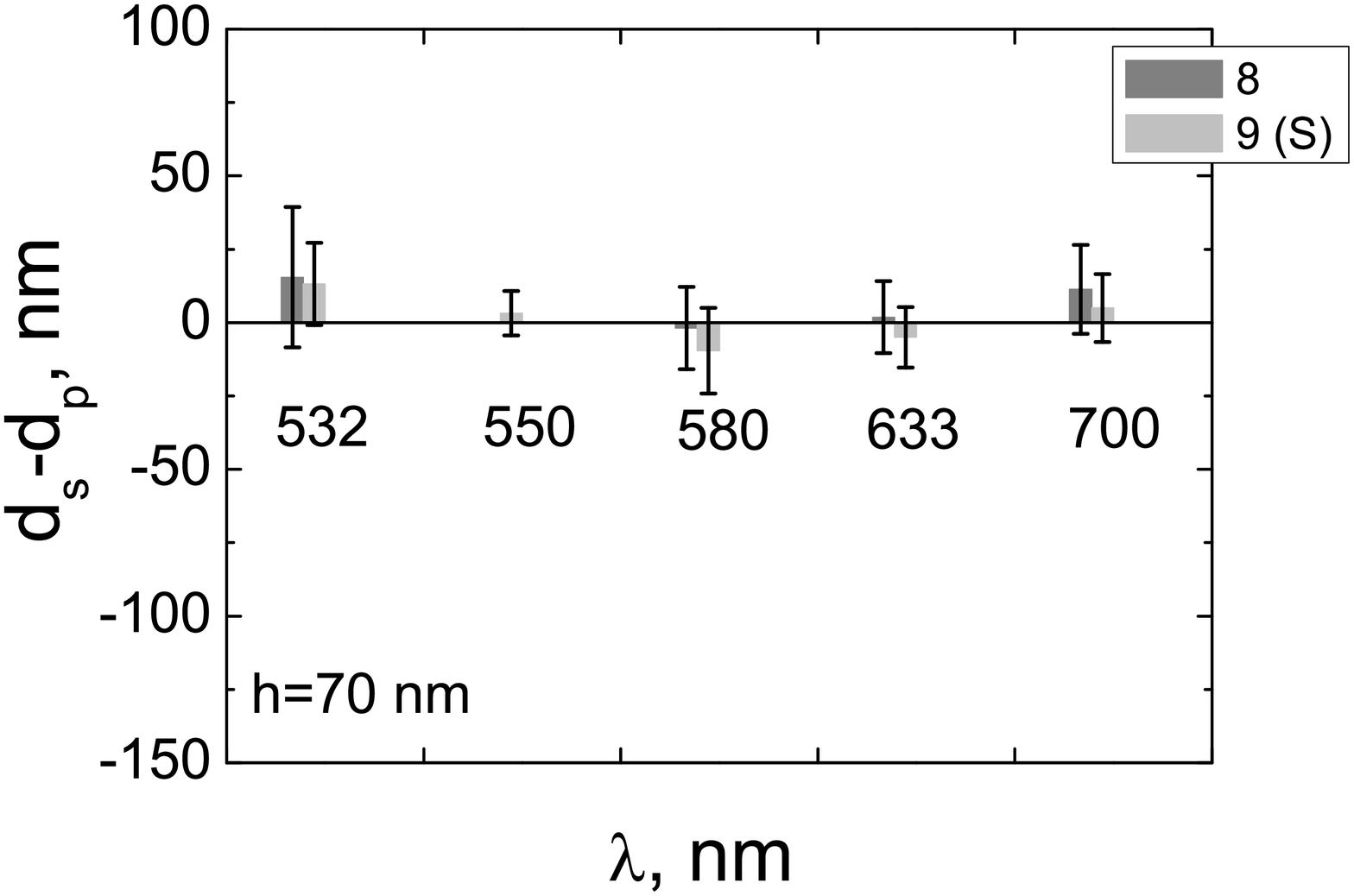} \text{(b)}
\centering{\includegraphics[scale=0.20]{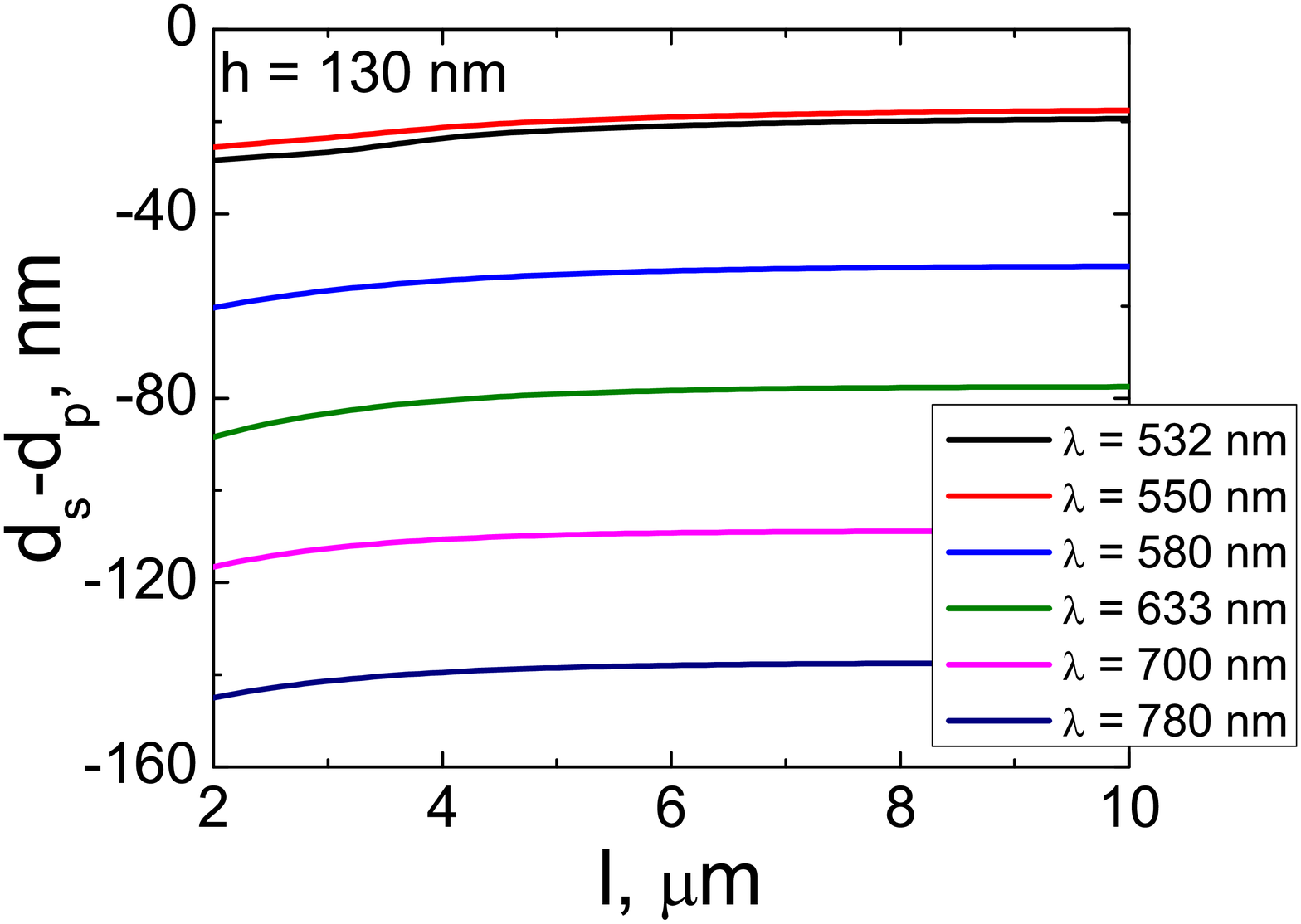}\text{(c)}}
\caption{Difference in the peak positions $d_s-d_p$ of the photocurrent's derivative vs wavelength $\lambda$ for various Au samples of $130$ (a) and $70$ (b) nm thickness. (T) means tempered samples ($30$ sec. $400$C$^\circ$). (S) means a sample with a standing-alone structure to investigate a possible influence of the opposite wall on the knife-edge measurement. In this case other structures are missing (a,b). Theoretical difference of the peak positions of the photocurrent's derivative ($d_s-d_p$ ) versus slit width $l$ for Au sample ($h=130$ nm) at various wavelengths (c). }
\label{fig:reproduzierbarkeit}
\end{figure}

\subsection{Principle of the measurement}
The principle of the measurement is depicted in Fig. \ref{fig:exp_setup}. The experiments are performed using a highly focused linearly polarized TEM$_{00}$-mode. We investigate two polarization directions of the incoming beam relative to the knife-edge (in the $x$-$y$ plane). At first the electric field is oriented perpendicularly (s-polarization) and then parallel (p-polarization) to the wall of the knife-edge (see Fig \ref{fig:exp_setup} (a), (b)). The investigated laser beam is blocked stepwise by the edge of an opaque metal stripe that is building a single knife-edge. The photocurrent generated inside the photodiode is recorded for each sample position $x_0$ (see Fig. \ref{fig:exp_setup} (c)). It is proportional to the power $P$ of the diffracted field detected by the photodiode
\begin{equation}
 P=\int _{-\infty}^0 S_z\left(x,x_0 \right)\mathrm{d}x,
\label{eq:knife}
\end{equation}
where $S_z$ is the $z$-component of the Poynting vector of the field at the photodiode. In the conventional knife-edge method the derivative of the photocurrent curve with respect to the sample position $x_0$ (see Fig. \ref{fig:exp_setup} (d)) reconstructs a beam projection on one axis \cite{JAArn71,AHFire77}, so the width and position of the projection can be determined.  

\begin{figure}[t!]
\centering
\includegraphics[scale=0.2]{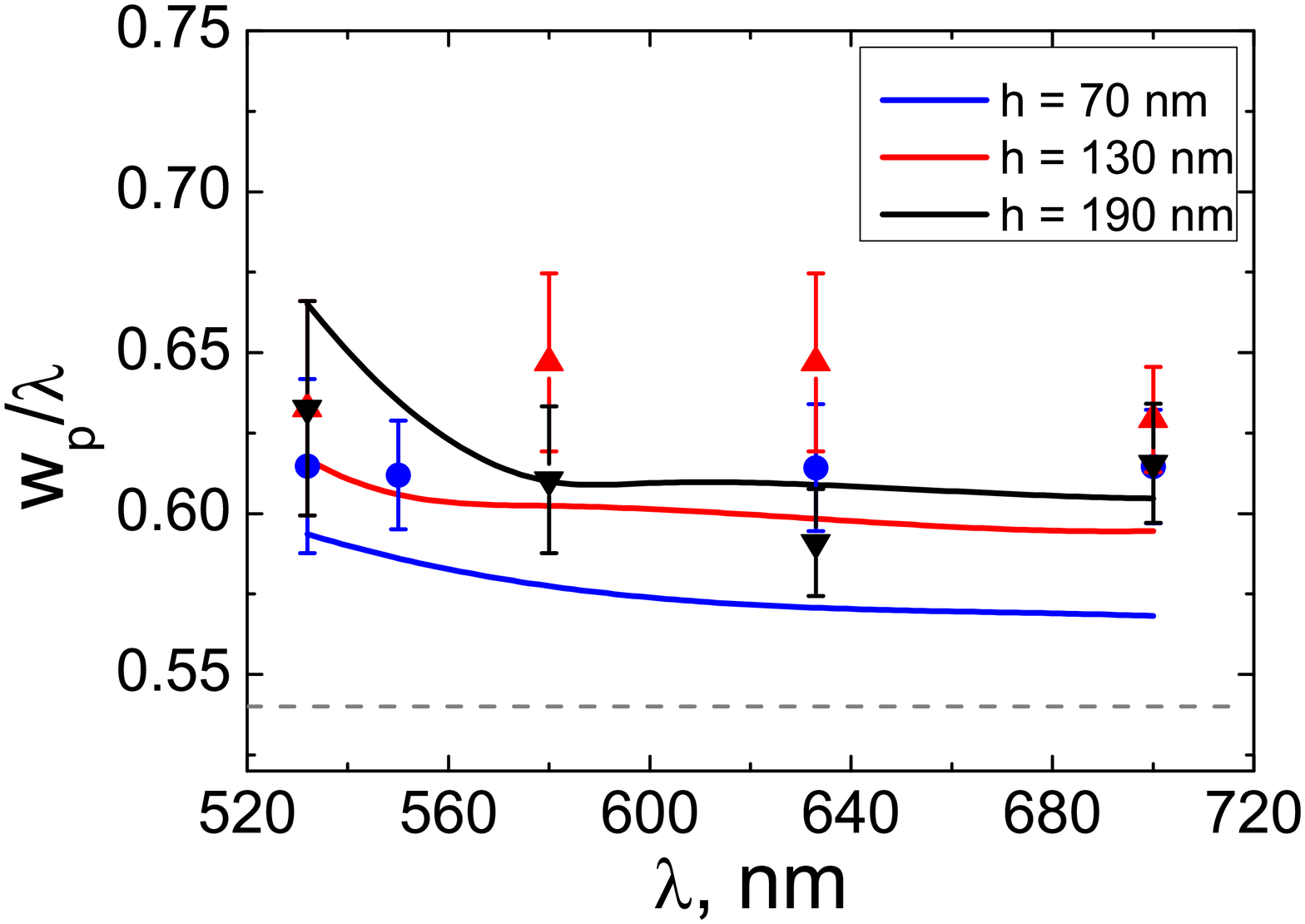}\text{(a)}
\includegraphics[scale=0.2]{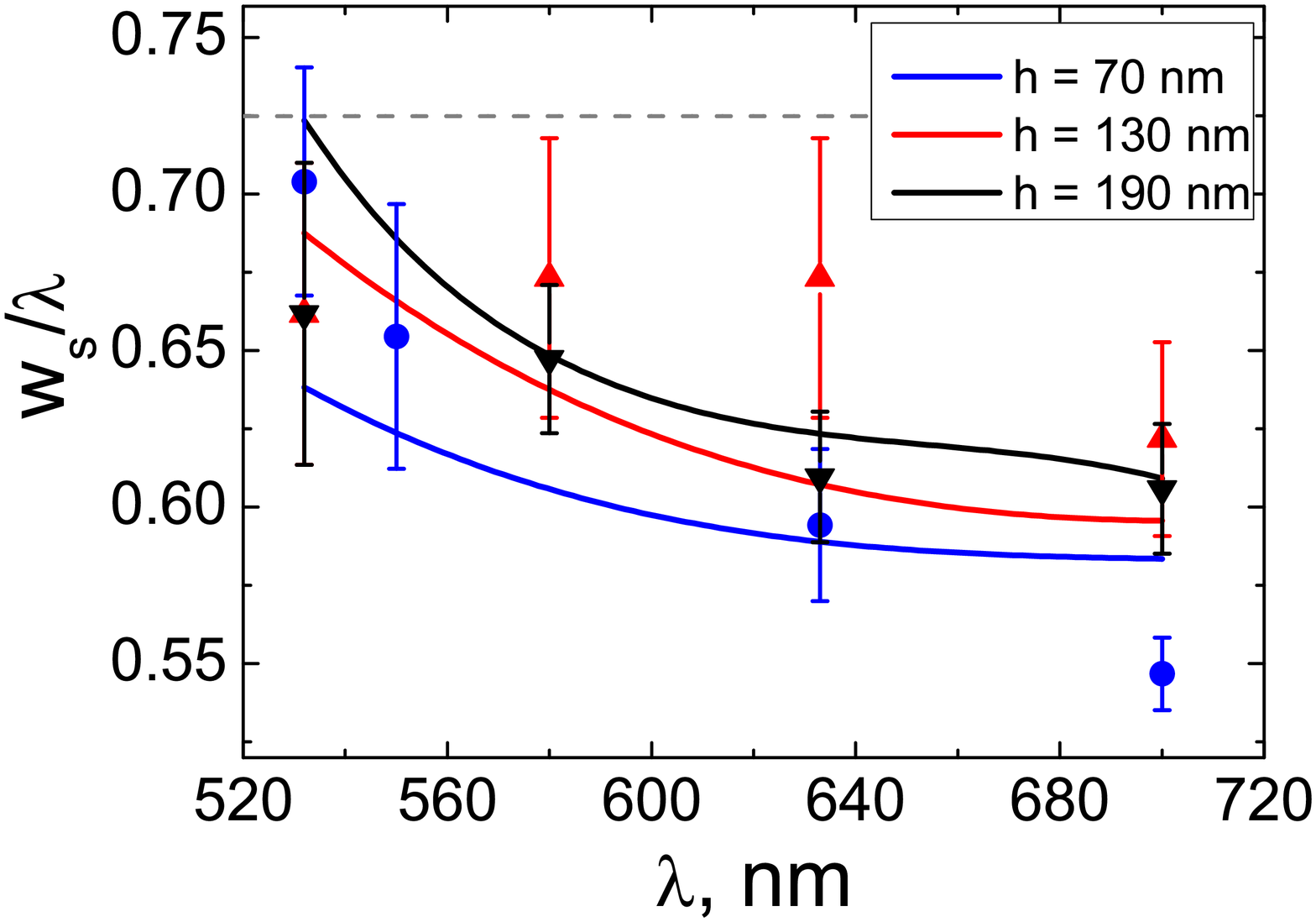} \text{(b)}  
\includegraphics[scale=0.2]{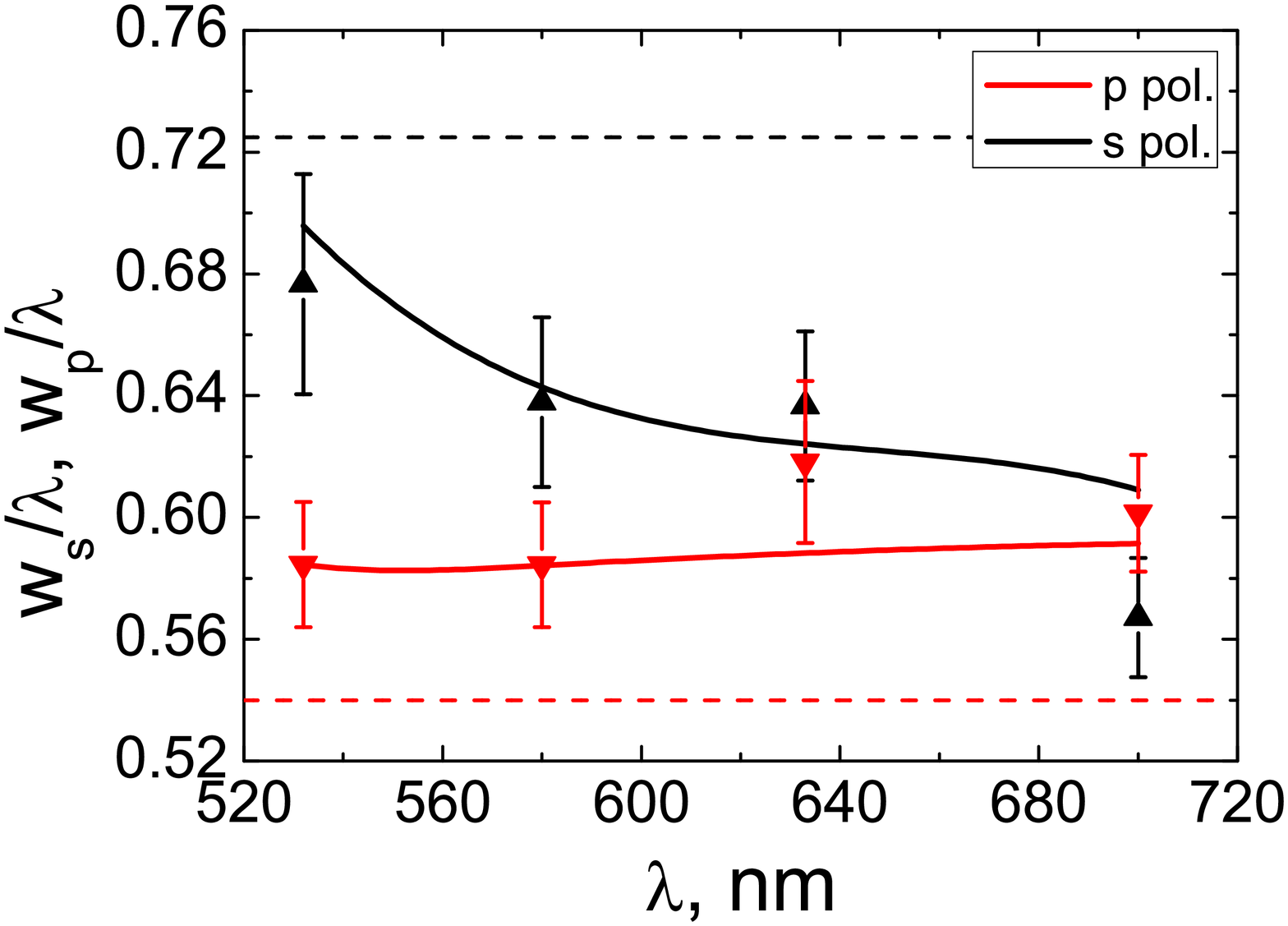}\text{(c)}
\includegraphics[scale=0.2]{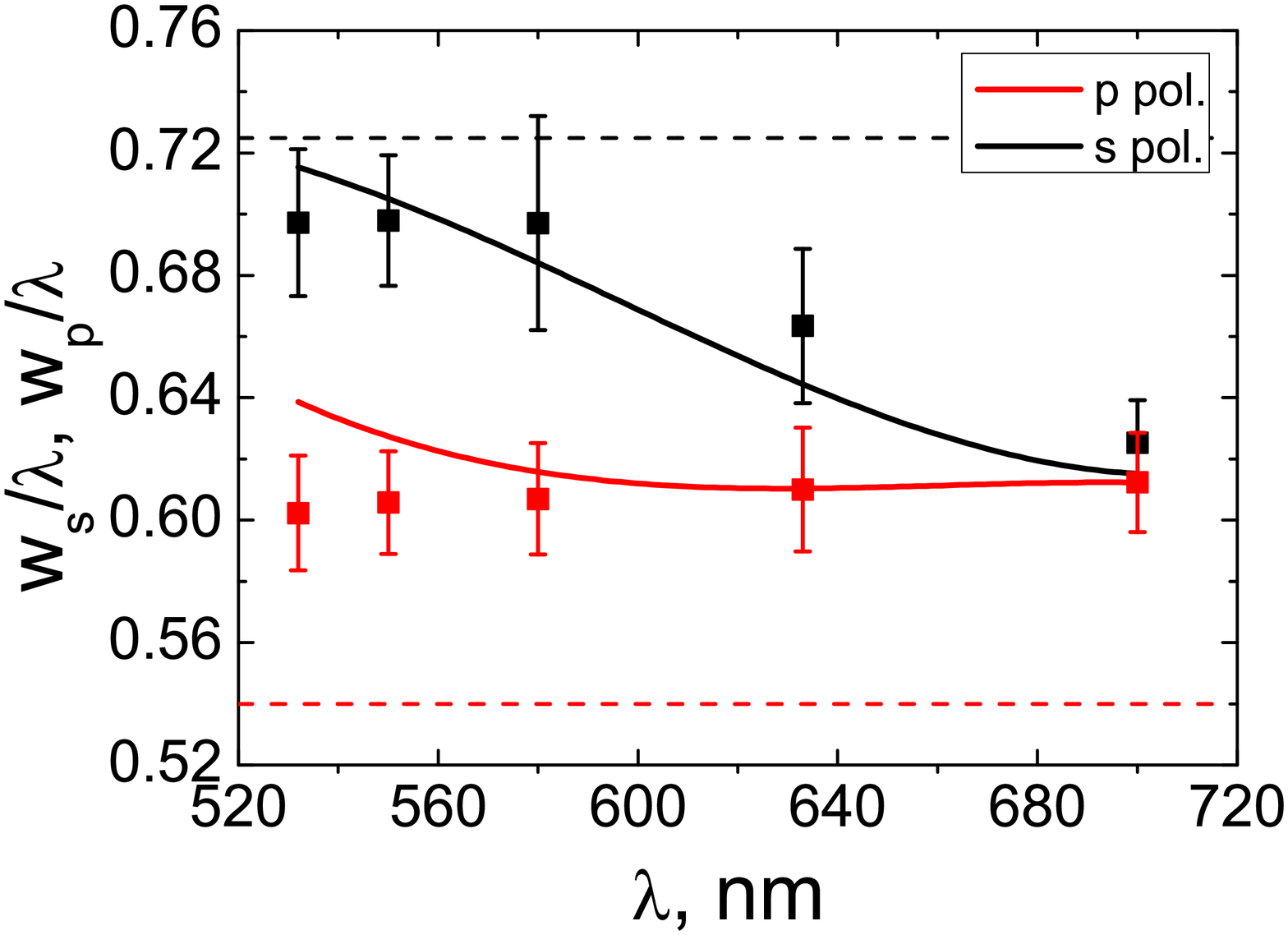}\text{(d)}
\caption{Dependence of the ratios $w_p/\lambda$ (a), $w_s/\lambda$ (b) on the wavelength $\lambda$ for Au samples of varying thickness. Dependence of the ratios $w_s/\lambda$, $w_s/\lambda$ on the on the wavelength $\lambda$ for Ni (c) and Ti (d) samples of comparable thickness of the opaque film $h=130$ nm. The colored curves represent results of numerical simulations, the colored points - of experimental measurements. Parameters of the model are $w_{0p}/\lambda=0.9w_{0s}/\lambda=0.55 $. The dashed lines represent the FWHM's of the squared electric field estimated from the Debye integrals \cite{BRic59}. }
\label{fig:exp_res2}
\end{figure}
Since the measurements with s- and p-polarized beams are performed one after another, a thermal drift of the sample between measurements for each polarization could introduce a systematic error. Therefore we profile the beam by two adjacent knife-edges, so the effects of sample drifts between measurements with different polarizations cancel out (see Fig. \ref{fig:exp_setup} (c)). After that the intensity of the beam projection is reconstructed from the recorded photocurrent and both the position and the width of the beam are evaluated. Two parameters $d_s$ and $d_p$ define the distance between the peaks of the reconstructed beam profiles (see Fig. \ref{fig:exp_setup} (d)). In the conventional knife-edge method no polarization effects are present, so the distance between the peaks for s- and p-polarized beam projections is equal ($d_s=d_p$). Thus, a non zero value for $d_s-d_p$ indicates the presence of the polarization effect. 

Measurements for each polarization are performed at $22$ different positions of the knife-edge. Each scan line in $x$-direction consists of $400$ steps with a step size of $10$ nm. For each position of the edge $10$ data points are measured and averaged to improve the signal-to-noise ratio. The photocurrent data was additionally filtered using a Savitzky-Golay smoothing algorithm (11 points) before the photocurrent curve was differentiated.

Most scans are performed at knife-edges with a width of approx. $2$ $\mu$m. Measurements on wider knife-edges (approximately $3$ and $4$ $\mu$m) do not result in any quantitative change within the experimental accuracy. Hence, in what follows we solely discuss results obtained for the $2$ $\mu$m structures.

During the measurement the relative position of the sample is automatically calibrated every $15$ min by the determination of the position of a $500$ nm hole. The hole is etched into one of the metal structures which is situated some microns away from the knife-edges. The detection of the hole position is estimated by the maximum of transmission through it. Thus long term stability is obtained and the actual sample position is always determined with a precision better than $50$ nm.

\subsection{Experimental results}
We start our discussion with experimental results for a Au knife-edge with a thickness (or height) of $h = 130$ nm. The beam profiling results $d_s$ and $d_p$ are presented in Fig. \ref{fig:exp_res} (a) for two polarization states ($s$ and $p$) and six particular wavelengths. In the conventional knife-edge theory, the zero crossing of the second derivative of the photocurrent curve coincides with the position $x$ in the photocurrent curve, where half of the maximum photocurrent is reached. It physically indicates the position of the knife's edge \cite{JAArn71,AHFire77}. So, the total distances between the peaks of the reconstructed profiles $d_s$ and $d_p$ can be interpreted intuitively as the width of the single stripe if no polarization effects are present.

As one can see in Fig. \ref{fig:exp_res} (a), the optical beam profiling reveals the polarization and wavelength dependence of the derived values $d_s$ and $d_p$. It should be noted that the obtained knife-edge width for the p-polarized beam remains almost constant as the wavelength changes, although it is actually about $100$ nm larger than the value determined by the SEM. Concerning s-polarization the knife-edge width $d_s$ depends on the wavelength and monotonically decreases as the wavelength increases. To characterize the polarization effect on the measurement we plot the value $d_s-d_p$ which also depends on the wavelength (see Fig. \ref{fig:exp_res} (b)).

Experimental data were obtained for Au samples of different thicknesses. The results for the relative shifts $d_s-d_p$ for Au are presented in Fig. \ref{fig:exp_res} (b) and (c). The profiling results of Au samples for different thickness shows a zero shift $d_s-d_p$ at approx. $\lambda=550$ nm for $h=100$ and $130$ nm and also for the whole wavelength range investigated at $h=70$ nm. Additionally, samples made from Ni and Ti were investigated. The results for Ni and Ti are very similar to each other (see Fig. \ref{fig:exp_res} (d)). As compared with Au of $h=130$ nm (see Fig. \ref{fig:exp_res} (b)) no zero shift was observed for Ni and Ti samples in the investigated wavelength range. Thus, the polarization effect depends not only on the thickness but also on the material of the knife-edge. Due to the roughness of the knife-edge and the photodiode surface (see Fig. \ref{fig:sample} (b)) the method is position dependent, what is the main source of systematical errors. The theoretical curves, which were obtained using theory developed in the next part of the paper, are also presented in the Fig. \ref{fig:exp_res} (b)-(d). Slight discrepancies between the experimental and theoretical results are due to the fact that in the experiment the edge position was determined as a point of maximum derivative of the probe current. Furthermore, the sample in the experiment had a non-zero slope angle (see Fig. \ref{fig:sample} (b)). Our numerical simulations reveal a slight asymmetry in the reconstructed beam profile (see discussion in the next section). So, the position of the maximum of the photocurrent
derivative does not coincide with the position of the half level of the photocurrent curve. However, if many measurements are averaged to reduce the influence of noise the curves become more inversion symmetric. Points of maximum derivative and those of half transmission start to merge. In fact, if we add noise to the numerical results and average afterwards the agreement between experimental and numerical results is even improved (see Fig \ref{fig:exp_res} (b)). We see a good qualitative agreement of our experimental and numerical findings for all samples. Only at a wavelength of $\lambda = 532$ (see Fig. \ref{fig:exp_res} (c)) and a knife-edge thickness of $h=100$ nm and $130$ nm we observe stronger discrepancies.

To check the reproducibility of the results several Au samples were produced. The relative shifts obtained in the measurements are presented in Fig. \ref{fig:reproduzierbarkeit} (a) for $h=130$ nm (sample No. 1-7) and (b) for $h=70$ nm (sample No. 8, 9). The results on sample No. 1 were already shown in Fig. \ref{fig:exp_res} (b). Samples No. 1-3 were fabricated using the same fabrication steps. Furthermore we investigated the dependency of the measured relative shift on the recrystallization of the knife-edge material (Au) by additionally tempering  (for $30$ s. at $400$ C$^\circ$) the sample after fabrication. By tempering the grainy metal surface becomes smooth and the surface roughness decreases. The results for the tempered Au samples are shown in Fig. \ref{fig:reproduzierbarkeit} (a) (sample No. 4-6). In Fig. \ref{fig:reproduzierbarkeit} (a) one can see that by tempering the knife-edges the relative shifts ($d_s-d_p$) derived from the measurements are slightly decreased for short wavelengths, although, the overall qualitative wavelength dependency is preserved (compare samples 4-6 with 1-3). The agreement between the experiment and the theory is slightly better for the tempered sample of the thickness $h=130$ nm, compare with Fig. \ref{fig:exp_res} (b). 

The aforementioned samples have a periodical geometry (see Fig. \ref{fig:sample} (a)). Therefore, we also check the dependence of the experimental results on the presence of an adjacent knife-edge. For this purpose single non-tempered knife-edges were fabricated (sample No. 7 ($h=130$ nm), see Fig. \ref{fig:reproduzierbarkeit} (a) and No. 9 ($h=70$ nm), see Fig.\ref{fig:reproduzierbarkeit} (b)). The measurements on these samples were performed under comparable conditions as described before. The results for these samples are similar to the results obtained with non-tempered periodically placed knife-edges and the differences were found to be in the range of the experimental errors. Such outcome is expected in our theoretical model, see next Section. The theoretical curves reveal changes of the same degree as in the experiment (see Fig.\ref{fig:reproduzierbarkeit} (c)).

As one can see measurement results depend on many parameters including the surface quality and roughness of the knife-edge. Therefore it is important to perform measurements of the relative shift at different positions of the knife-edge, which was done in all measurements performed here.

Following our original intention we now derive the beam width from our experimental data (see Fig. \ref{fig:exp_res2}). We explicitly do this for the sample No. 1 (Au, h=130 nm). To this end we measure the FWHM of the first derivative of the photo-current signal peak (see Fig. \ref{fig:exp_setup} (d)).  As the beam-width is expected to depend on the wavelength linearly, we investigate the ratios $w_s/\lambda$ and $w_p/\lambda$, which should be constant. As we see from the experimental data, for samples of thickness $h=130$ nm, the measured FWHM remains indeed nearly constant for p-polarization (see Fig. \ref{fig:exp_res2} (a)) but decreases for s-polarization with growing wavelength (see Fig. \ref{fig:exp_res2} (b)). The theoretical predictions are also shown in the pictures. With increasing thickness $h$ of the knife-edge the width of the reconstructed profile increases (see Fig. \ref{fig:exp_res2}). This accounts to the theoretical prediction, that the slit acts as a complex spatial frequency filter which damps modes with higher spatial frequency. In order to accurately measure the beam, the spatial spectrum of the signal has to be preserved, while the beam is blocked by the knife-edge. Obviously this can be accomplished by a knife-edge of small enough thickness. Next, as the wavelength increases, the reconstructed width of the beam slowly approaches its theoretical value. However, the reconstruction of the p-polarized beam reveals, that the reconstructed width of the beam projection remains larger than its real value over the investigated wavelength range. Due to the rather complicated interplay of the underlying effects, we can only suggest, that this is caused by the remaining power-flow through the wall of the knife (see next section for more details). Nevertheless the agreement between the experiment and the simulations is good, however, the FWHM values obtained from the Debye integrals \cite{BRic59} and reconstructed experimentally are different (see Fig. \ref{fig:exp_res2}).  In what follows we develop a theoretical model to gain some insight into the complex dynamics of light-matter interaction at a knife-edge.  

\section{Theoretical model}
\subsection{Realistic knife-edge versus perfect knife-edge}
We start the theoretical discussion with a brief reminder of the knife-edge method basics. In the original work \cite{JAArn71} the following assumptions were made: the incident beam is paraxial, the knife-edge is made from a perfect conductor and no losses are present. For sake of simplicity we consider a two-dimensional situation. The incident field in the spectral domain is represented as plane waves with amplitudes $S\left(k_x \right)$ traveling at different angles $\alpha = \arcsin k_x/k$. Here $k=\omega/c_0$ is the wave vector, with $k_x$ and $k_z$ being the transverse and longitudinal components of the wave vector, $\omega$ is the frequency and $c_0$ is the speed of light in vacuum. In the first part of Figure \ref{fig:theor_setup} (a) a single plane wave component $\mathbf{k}=\left(k_x,k_z \right)=k\left(\pm  \sin \alpha, -\cos \alpha \right)$ of the spatial spectrum $S\left( k_x \right)$ of the Gaussian beam is shown. Three possible interaction scenarios are shown in different colors. First, the plane wave part, represented by the red rays is blocked by the knife-edge and reflected from the metal surface. 
Second, the partial ray shown in green is not affected by the knife-edge and is properly detected. Furthermore, the wave part shown in blue experiences reflection ($r=1$) from the wall of the knife-edge, however its amplitude remains the same and is therefore also detected properly. 

Next, we introduce a finite conductivity of the knife-edge but still consider paraxial beams, i.e. reflection $r \approx 1$ (see Fig. \ref{fig:theor_setup} (a)). At the side wall the boundary conditions for s- and p-polarizations are $E_x = \epsilon E_x^{(2)}$ and $E_y = E_y^{(2)}$, where superscript $2$ denotes the electric field component in the knife-edge and $\epsilon$ is a dielectric constant of the knife-edge material. The fields in the knife-edge are decaying exponentially as $\exp \left(-\kappa kx \right)$, where $\kappa= \mathrm{Im} \epsilon ^{1/2}$. They still contribute to the total detected power $T$ as an additional term $\exp\left(- \frac{x_0^2}{d_0^2}\right)\frac{\mathrm{Re}\upsilon}{\kappa k}$, where $d_0$ is the FWHM of the intensity $\left|E\right|^2$ of the scalar beam, $\upsilon = \epsilon ^{-1}$ for a TM incoming field and $\upsilon = 1$ for the TE field. The differentiation of the transmission gives us
\begin{equation}
\frac{\partial T}{\partial x_0}= -\exp \left(-\frac{x_0^2}{d_0^2} \right)-\frac{2x_0}{d_0^2}\exp \left(-\frac{x_0^2}{d_0^2} \right)\frac{\mathrm{Re}\upsilon}{\kappa k}
\label{eq:dTrans1}
\end{equation}
and can be interpreted as the beam projection (compare with Section 2). The second term leads to an asymmetric profile and results in a shift of the projection maximum, which therefore does not longer correspond to the half maximum of the transmission curve $T$. The parameter $\upsilon$ depends on the polarization relative to the edge. Thus, the shift of the maximum in the projection curve is polarization dependent in this simple model, which leads, in general, to a non-zero relative shift $d_s-d_p$.

As a next step towards a realistic knife-edge, we introduce an angle dependent reflection $r\left(k_x \right)$ from the side wall of the knife-edge (see Fig. \ref{fig:theor_setup} (a)). For s- and p-polarization the amplitudes of the reflected plane wave parts are $E_{ref}=E_{inc}r\exp\left(\mathrm{i}\delta \right)$, here $\delta$ is an angle-dependent relative phase of the reflected plane wave. In the paraxial case it holds: $r \approx 1 - C\alpha$ and $\delta \approx D \alpha$, where $C$ and $D$ are different for s- ($\bot$) and p- ($||$) polarization. $C$ and $D$ are obtained from Fresnel's equations and $k_x=k\alpha$. Thus, the reflected part of the plane wave can be written as $E_{ref}= \exp\left(-\mathrm{i}k x \alpha + \ln r - \mathrm{i}\delta \right)$. Next we use $\ln r \approx -C \alpha$. The single plane wave component affected by the reflection can be written as $\exp\left(\mathrm{i}kx \alpha \right)+\exp\left[-\left(\mathrm{i}k x +C  + \mathrm{i}D\right)\alpha \right]$. In the perfect case (Fig. \ref{fig:theor_setup} (a)) the spatial spectra of the signal blocked by the knife-edge at the photodiode is $S_{sig}\left(k \alpha \right)=S\left(k\alpha\right) \exp \left(-\mathrm{i}kx_0 \alpha \right)$. However due to the reflection, the plane wave component $-k_x$ is replaced by the aforementioned expression. After some mathematical operations it can be revealed that the reflection from the wall of the knife-edge introduces further modifications to the signal integrated by the detector. The signal experiences a polarization dependent shift $\Delta x = -D/(2k)$, it is spatially filtered and the shape of the spectral profile is deformed.

We note, that a possible solution to make the method less polarization dependent is the usage of a knife-edge with a rough surface, so the reflection will be polarization independent. For instance, alloys with unstable binary phases (Zn and Au) were used as knife-edge materials in previous investigations successfully \cite{RDorn03a,RDorn03b}.

\subsection{Field representations and eigenvalue problem}
\begin{figure}[t!]
\centering
\includegraphics[scale=0.2]{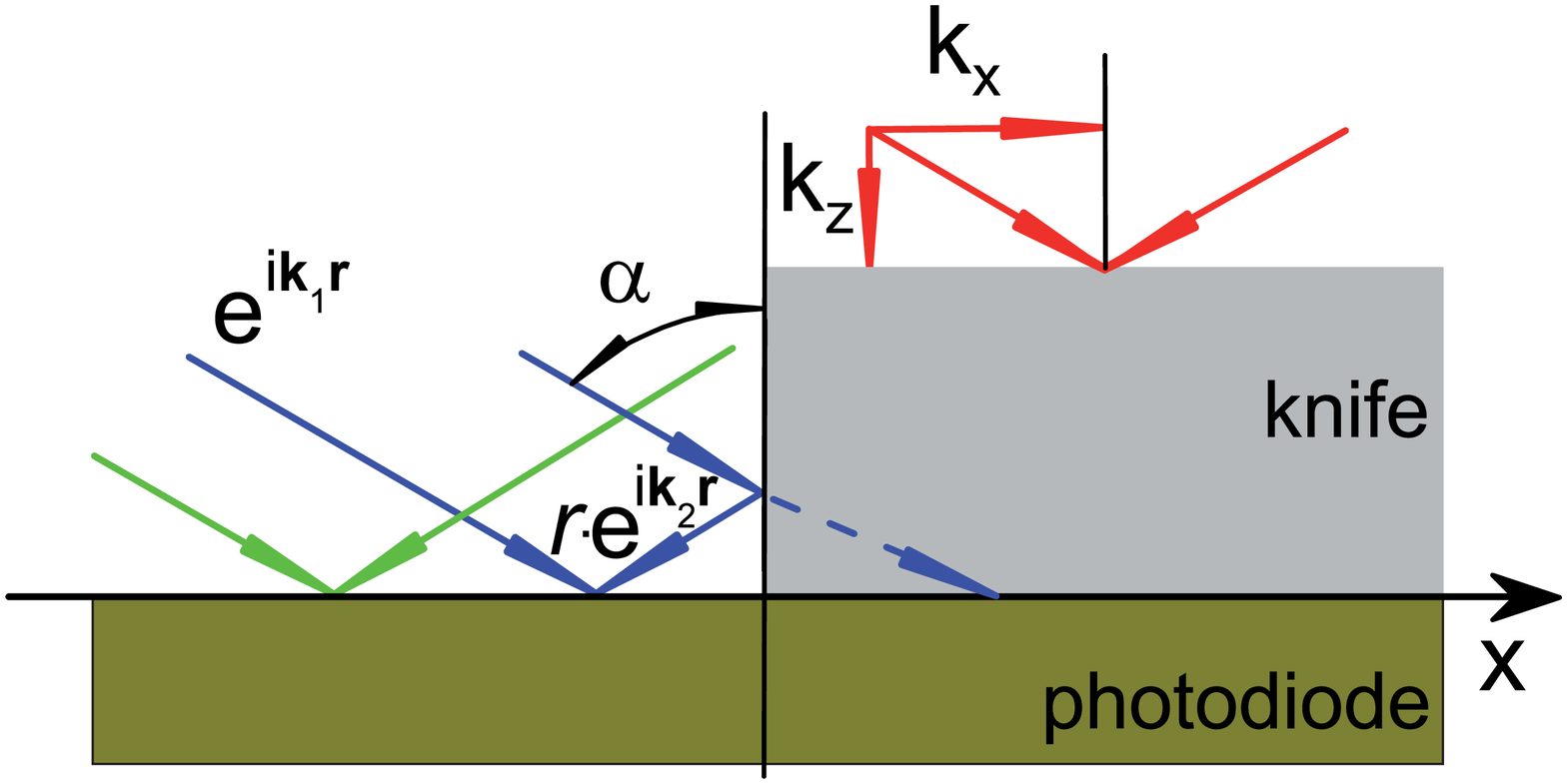}\text{(a)} 
\includegraphics[scale=0.2]{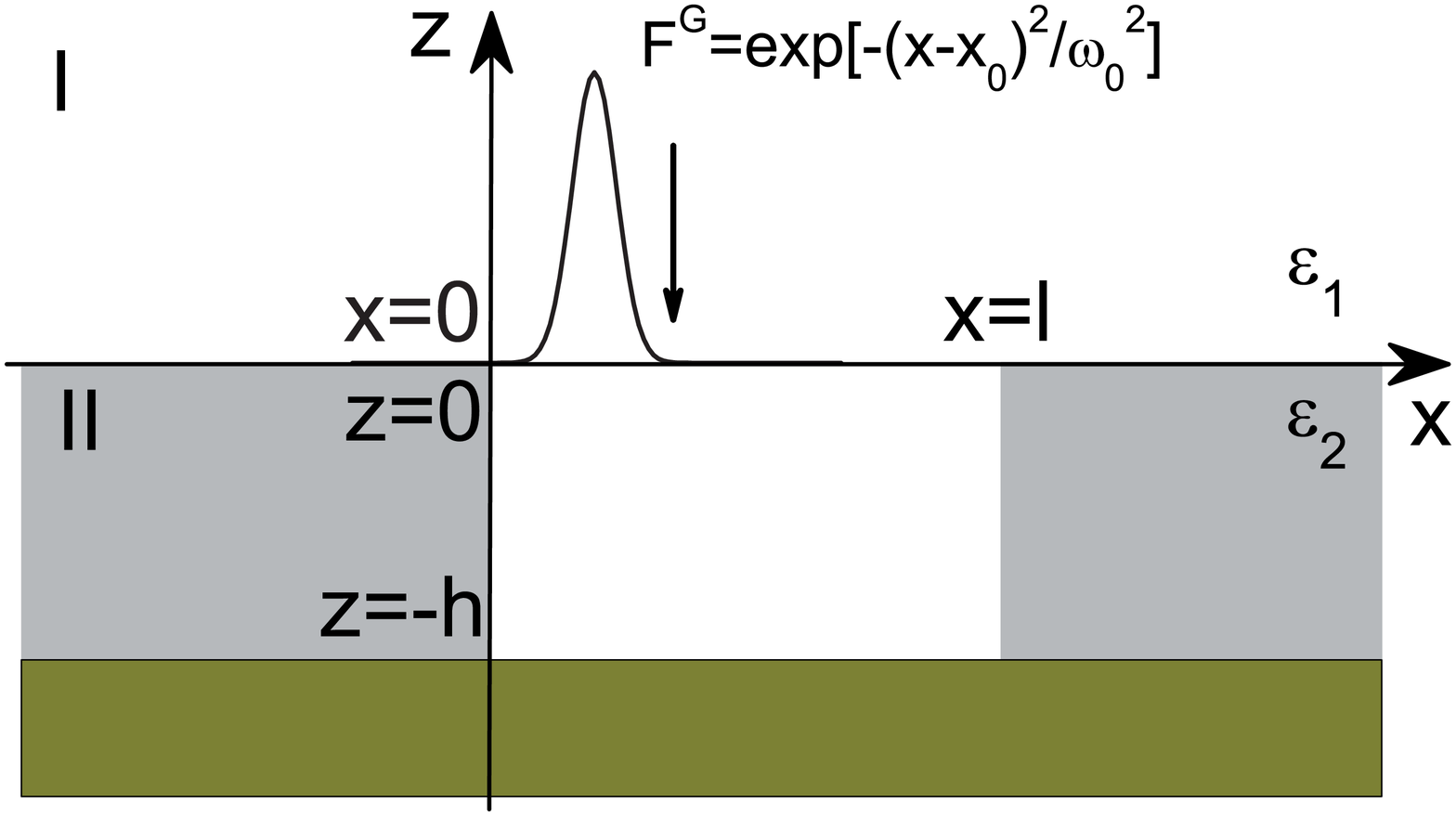}\text{(b)}
\caption{Schematic depiction of a single plane wave component $\mathbf{k}=\left(\pm k_x,k_z\right)$ impinging on a knife-edge. Here $\mathbf{k}_1=\left(k_x,k_z \right)$, $\mathbf{k}_2=\left(-k_x,k_z \right)$ and $\mathrm{r}=\left(x,z\right)$ (a). Sketch of the considered structure (b).}
\label{fig:theor_setup}
\end{figure}
The theoretical consideration of the previous subsection, though it is helpful for the understanding of the underlying principles, has a number of limitations. In this section we proceed with the development of an exact theoretical model, which considers the experimental situation exactly: vector fields and plasmonic modes, reflection from the photodiode and exact boundary conditions will be also included. We start with a formulation of an eigenvalue problem.

We consider the structure represented in Fig. \ref{fig:theor_setup} (b), where two adjacent knife-edges build a wide ($l>\lambda$) slit. The space is divided into three regions, of which region I (half space $z>0$) and III (half space $z<-h$) are assumed to be dielectric and homogeneous with dielectric constants $\epsilon_{1}$ and $\epsilon _{3}$, respectively. The intermediate region II ($0\geq z \geq -h$) of thickness $h$ consists of an opaque (i.e. the skin-effect depth is smaller than the height $h$) material (for $x\leq0$ and $x>l$) with a dielectric constant $\epsilon _2$ and empty space (for $l>x>0$), with the same dielectric constant as region I, which we put to unity $\epsilon_1=1$. The considered structure is piece-wise homogeneous and in each homogeneous part of the structure solutions of the Helmholtz equation describe the field propagation. A continuous solution up to first order of the derivatives is derived by combining solutions of homogeneous parts using appropriate boundary conditions. In the general case, in the two-dimensional waveguide structure with ohmic losses in the walls the coupling between transverse electric (TE) and transverse magnetic (TM) modes was reported \cite{AEKar55}. So, as a further simplification, we restrict our consideration to a planar approximation, assuming that the incident field does not change in $y$ direction, so the Helmholtz equation in each homogeneous part can be written in it's two-dimensional form
\begin{align}
\left(\frac{\partial ^2}{\partial x^2}+\frac{\partial ^2}{\partial z^2}+k_{\epsilon}^2 \right) \left \{ \begin{array}{ll}
	 \mathbf{E}\left(x,z \right) \\
	\mathbf{H}\left(x,z \right)
	\end{array} \right \} = 0, \quad k_{\epsilon} = \frac{\omega\sqrt{\epsilon}}{c_0},
\label{eq:Helmholtz}
\end{align}
$\epsilon$ denotes the respective dielectric constant. The solution of  equation (\ref{eq:Helmholtz}) consists of two independent classes: transverse electric and transverse magnetic modes, propagating in $z$ direction.

We start our consideration with TE modes of the Region II. The solution of the two-dimensional Helmholtz equation (\ref{eq:Helmholtz}) for the electric field $\mathbf{E}$ in this case has only one nonvanishing component of the electric field $E_y$ parallel (p-polarization) to the slit walls
\begin{equation}
H_x = - \frac{\beta _ec_0}{\omega Z_0}E_y, \ H_z=\frac{\mathrm{i}c_0}{\omega Z_0}\frac{\partial E_y}{\partial x}, \ E_x = E_z = H_y = 0.
\label{eq:TEmode}
\end{equation}
Here $\beta _e$ is the effective propagation constant of the electric field mode in the empty part of region II (for the sake of brevity we will call it slit) and $Z_0$ is the vacuum impedance.

Next, the transverse magnetic (TM) modes of the metal slit have one nonvanishing component of the magnetic field $H_y$ and electric field components are perpendicular (s-polarization) to the walls of the knife-edges
\begin{equation}
E_x =  \frac{\beta _m}{\omega \epsilon _0 \epsilon \left(x \right)}H_y, \ E_z=-\frac{\mathrm{i}}{\omega \epsilon_0 \epsilon \left(x \right)}\frac{\partial H_y}{\partial x}, \ H_x  = H_z = E_y = 0.
\label{eq:TMmode}
\end{equation}
Here $\beta _m$ is the propagation constant of the magnetic field mode in the slit and $\epsilon(x)=1$ for $0<x<l$ and $\epsilon(x)=\epsilon_2$ elsewhere.

In general, the eigenmodes of the metal-insulator-metal waveguide can be divided into the two classes: the localized and nonlocalized solutions of Eq. \ref{eq:Helmholtz}, see Ref. \cite{BStur07}. The first class has a discrete spectrum of the eigenvalues and represents the field localized in the slit. The second class has a continuous spectrum of the eigenvalues and is necessary to ensure the boundary conditions on the metal surface $x\notin \left(0,l \right)$. 

However, when light penetrates the metal knife directly from top, the electric field will decay exponentially in the metal \cite{SEKoc09}. Therefore, the non-localized solutions can be neglected when calculating the transmission. We verified the validity of this further simplification numerically. Hence, we construct only localized eigenmodes of the three-piece region II, which consists of three functions: the first and the last one describes plane waves propagating into the wall, the second one represents an interference of two plane waves inside the slit. So the eigenfunctions of the Helmholtz equation $E _{y,\nu}$ and $H _{y,\nu}$ have the following form in region II
\begin{align}
\left \{
\begin{array}{l}
E_{y,\nu} \left( x \right) \\
H_{y,\nu} \left( x \right)	
\end{array}  \right \}
 = e^{\pm \mathrm{i}\beta _{(e,m)} z}
  \left \{ \begin{array}{ll}
	 C_1 e^{\gamma _2 x}, \quad \gamma _2 = \sqrt{\beta _{(e,m)}^2 -k^2 \epsilon _2},\ \text{if} \ x\in \left[-\infty, 0\right]  \\
	C_2 e^{\gamma _1 x} + C_3e^{-\gamma _1 x}, \quad \gamma _1 = \sqrt{\beta _{(e,m)}^2 -k^2\epsilon _1 },\ \text{if} \ x\in \left [0, l \right] \\
	 C_4 e^{-\gamma _2 \left( x - l \right)} , \quad \gamma _2 = \sqrt{\beta ^2_{(e,m)} -k^2 \epsilon _2},\ \text{if} \ x\in \left[l, \infty \right]
	\end{array} \right.  
\label{eq:TEansatz}
\end{align}
where $\nu > 0$ is an integer  number, enumerating the eigenfunctions. $\gamma _2=\gamma _2^{'}+\mathrm{i}\gamma _2^{''}$ is the transverse part of the complex wave vector in the slit wall, $\gamma _1=\gamma _1^{'}+\mathrm{i}\gamma _1^{''}$ is the transverse wave vector in the slit and $l$ is the slit width. Here, plus and minus signs account for forward and backward propagating waves in the slit. The constants $C_1$, $C_2$, $C_3$ and $C_4$ in Eq. (\ref{eq:TEansatz}) are eliminated by enforcing the continuity of either the $E_y$ and $H_z$ components for the TE polarization or the $H_y$ and $E_z$ components for the TM polarization. This procedure results in two characteristic equations
\begin{equation}
e^{2\gamma _1 l} = \left( \frac{\gamma _1 - \gamma _2}{\gamma _1 + \gamma _2}	\right)^2 \\ \text{for TE}, \quad e^{2\gamma _1 l} = \left( \frac{\gamma _1 \epsilon _2 - \gamma _2}{\gamma _1 \epsilon _2 + \gamma _2}	\right)^2 \\ \text{for TM}
	\label{eq:TEequa}
\end{equation}
from which the modal propagation constants $\beta _e$ and $\beta _m$ are found correspondingly, \cite{BStur07}. The electric field $E_{y,\nu}$ and the magnetic field $H_{y,\nu}$ of the eigenfunction in TE and TM cases are expressed as $\pi _{\nu}\left(x \right) \exp \left(\pm \mathrm{i} \beta _{(e,m)} z \right)$, where
\begin{align}
\pi _{\nu}\left(x \right) = \left(1 - (-1)^{\nu}e^{\gamma _1 l} \right) e^{\gamma _2 x}, \ \text{if} \ x\in \left[-\infty, 0\right]  \nonumber \\
\pi _{\nu}\left(x \right) = e^{\gamma _1 x} -(-1)^{\nu} e^{-\gamma _1 \left(x - l \right)}, \ \text{if} \ x\in \left [0, l \right]   \nonumber \\
\pi _{\nu}\left(x \right) = \left(e^{\gamma _1 l} -(-1)^{\nu}\right) e^{-\gamma _2 \left( x - l \right)} , \ \text{if} \ x\in \left[l, \infty \right],  
\label{eq:TEmode_exp}
\end{align}
where the mode index $\nu$ starts with $\nu =0$ for TM and $\nu = 1$ for TE waves. We see that the mathematical expressions for the fields in TE and TM cases are functionally the same with the differences being hidden in the eigenvalues $\gamma _1$ and $\gamma _2$, which are derived from different equations, see (\ref{eq:TEequa}). We note here, that the boundary conditions at the walls are satisfied by Eq. (\ref{eq:TEmode_exp}), so Eq. (\ref{eq:TEansatz}) represents a field structure in Region II. The effects considered in the previous subsection are automatically included into the exact model.

In general, the eigenvalues $\beta _{(e,m)}$ of equation (\ref{eq:TEequa}) are complex for complex dielectric constants. Therefore, propagating modes are attenuated in the slit. However, the eigenmodes with the modal number $\nu>\nu_{cr}$, where $\nu_{cr}$ is estimated from the relation $k^2\epsilon _r -\pi ^2 \nu_{cr}^2/l^2\leq 0$, are attenuated much faster (also for lossless materials). The last note here is that solutions of eigenvalue equation (\ref{eq:TEequa}) in the TM case are divided into two families. The first family consists of two TM eigenmodes describing symmetric and antisymmetric plasmonic modes propagating along the walls of the slit and having evanescent field components not only in the metal but also in the slit \cite{JADio06,PLal06}. The second family of modes consists of fields, which oscillate in transverse direction in the slit. All eigenfunctions are orthogonal in the sense

\begin{equation}
\int _{-\infty}^{\infty} E_{y(x),\nu}H_{x(y),\mu}dx = \delta _{\nu\mu}G_{\nu \nu},
\label{eq:orthogon}
\end{equation}
where $\delta _{\nu\mu}$ is the Kronecker delta and $G_{\nu \nu}$ is an overlap integral of the electric and magnetic transverse fields of the eigenmode (\ref{eq:TEmode_exp}), see Ref. \cite{SEKoc09} for more details.
However the function $G_{\nu \nu}$ is in general a complex-valued one, therefore, we normalize our eigenmodes by introducing a real-valued norma $N_{\nu}$, which we define as
\begin{equation}
	N_{\nu}^2 = \int _{-\infty}^{\infty}\upsilon(x) \pi _{\nu} \left(x \right) \pi _{\nu}^{*} \left( x\right) dx,
	\label{eq:Normadef}
\end{equation}
here, $\upsilon (x) = \epsilon(x)^{-1}$ for TM and $\upsilon (x) = 1$ for TE polarization. An asterisk denotes a complex conjugation. Those expressions are readily evaluated to
\begin{equation}
	N_{\nu}^2 = \frac{1+e^{2\gamma _1^{'}l} - (-1)^{\nu} 2 e^{\gamma _1^{'}l} \cos \gamma _1^{''}l}{\gamma _2^{'}}- (-1)^{\nu}\frac{2\upsilon e^{\gamma _1^{'}l} \sin \gamma _1^{''}l}{\gamma _1^{''}}+\frac{\upsilon e^{2\gamma _1^{'} l}-1}{ \gamma _1^{'}}.
	\label{eq:TEnorma}
\end{equation}
here, $\upsilon   = \epsilon_2$ for TM and $\upsilon  = 1$ for TE polarization. The normalization is, in general, not necessary, however, this procedure ensures a slightly better convergance in our numerics.

In what follows we will evaluate boundary conditions between the different regions I and II or II and III in Fourier space. Therefore we derive the Fourier transform of the eigenfunctions in the slit. The Fourier image of the eigenfunction $\pi _{\nu} \left(x \right)$ can be written as a sum of two images. The image $\Phi _{\nu}$ represents the field in the walls, and $\Psi _{\nu}$ - the field in the slit

\begin{align}
\Pi _{\nu}\left(k_x \right ) = & \Phi _{\nu}\left(k_x \right )+ \Psi _{\nu}\left(k_x \right ), \nonumber \\
	\Phi _{\nu}\left(k_x \right ) = & \left [1-\left( -1
	\right)^{\nu} e^{\gamma _1^{*}l} \right]\frac{\gamma _2^{*}+k_x\mathrm{i}-\left( -1
	\right)^{\nu} e^{-\mathrm{i}k_xl}\left(\gamma _2^{*}-k_x\mathrm{i} \right) }{\sqrt{2 \pi N_{\nu}}\left(\gamma ^{*2}_2+k_x^2 \right)},  \nonumber \\
	\Psi _{\nu}\left(k_x \right ) = & \frac{\left(\gamma _1^{*}+k_x\mathrm{i} \right)\left(e^{\left(\gamma _1^{*}-\mathrm{i}k_x \right)l}-1 \right)-\left( -1
	\right)^{\nu} \left(\gamma _1^{*}-k_x\mathrm{i} \right) \left(e^{\gamma _1^{*} l}-e^{-\mathrm{i}k_xl} \right)}{\sqrt{2 \pi N_{\nu}}\left(\gamma ^{*2}_1+k_x^2 \right)}.
	\label{eq:TE_fourier}
\end{align}
\begin{figure}[t!]
\centering
\includegraphics[scale=0.3]{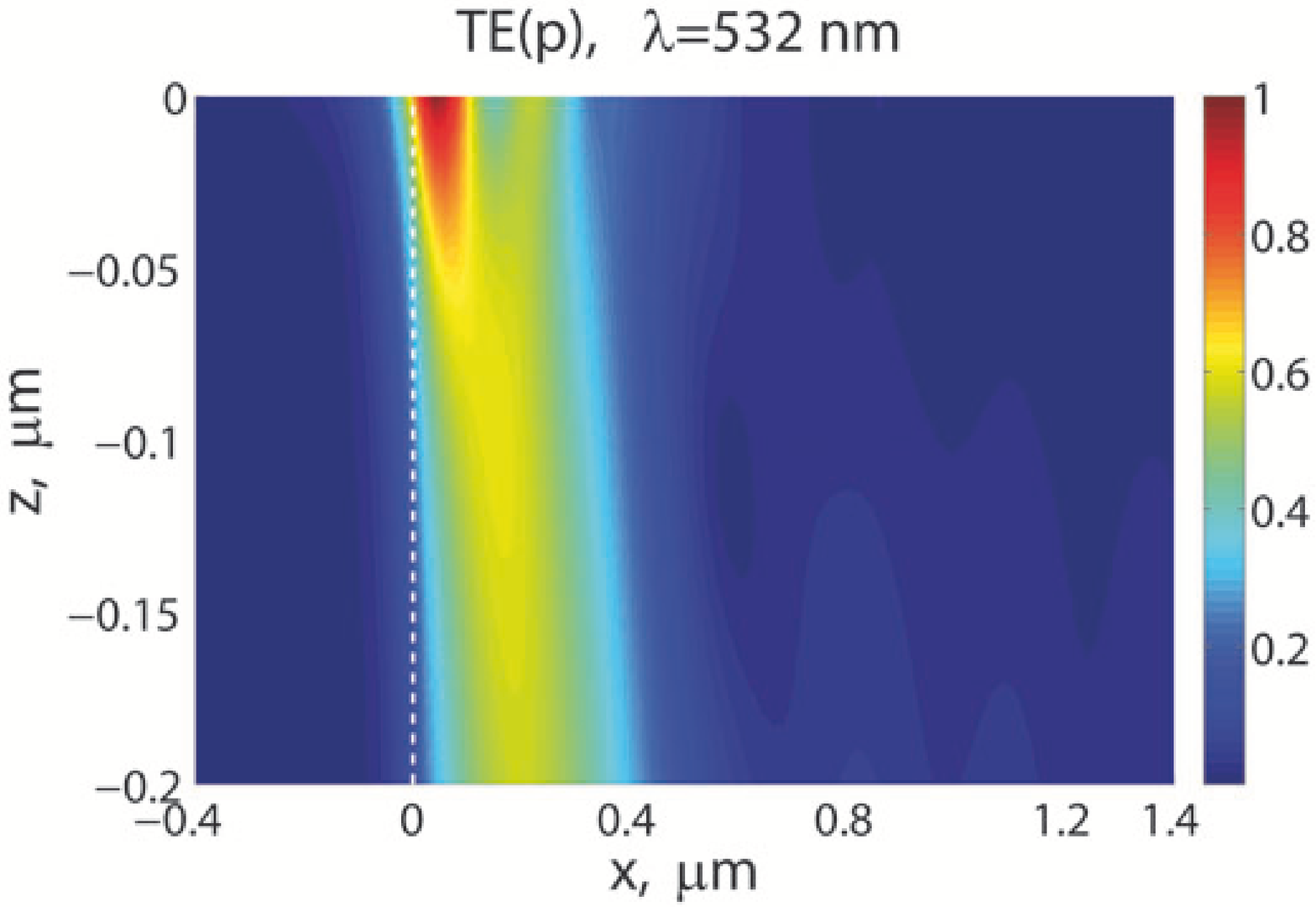} \text{(a)}
\includegraphics[scale=0.3]{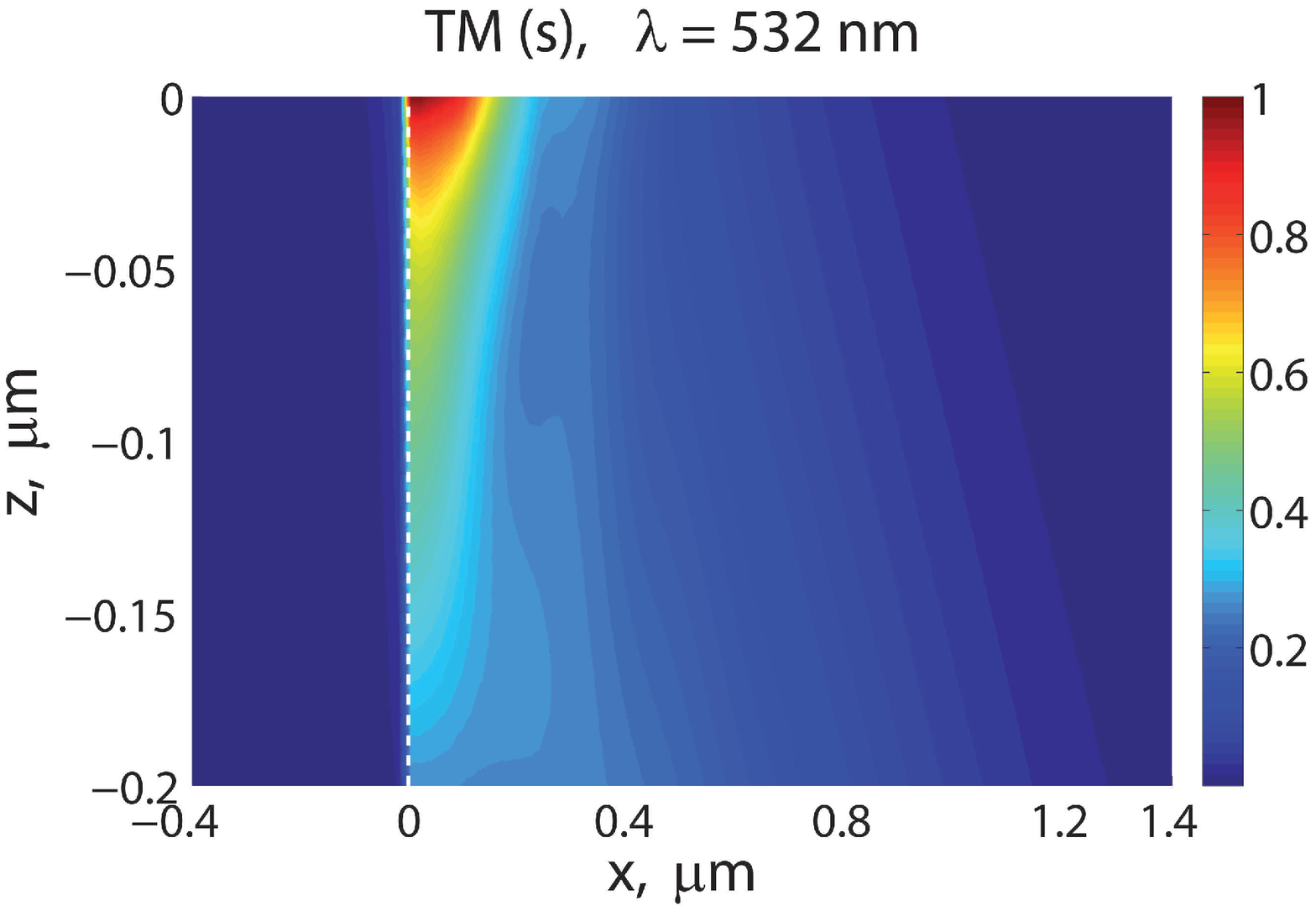} \text{(b)}
\includegraphics[scale=0.3]{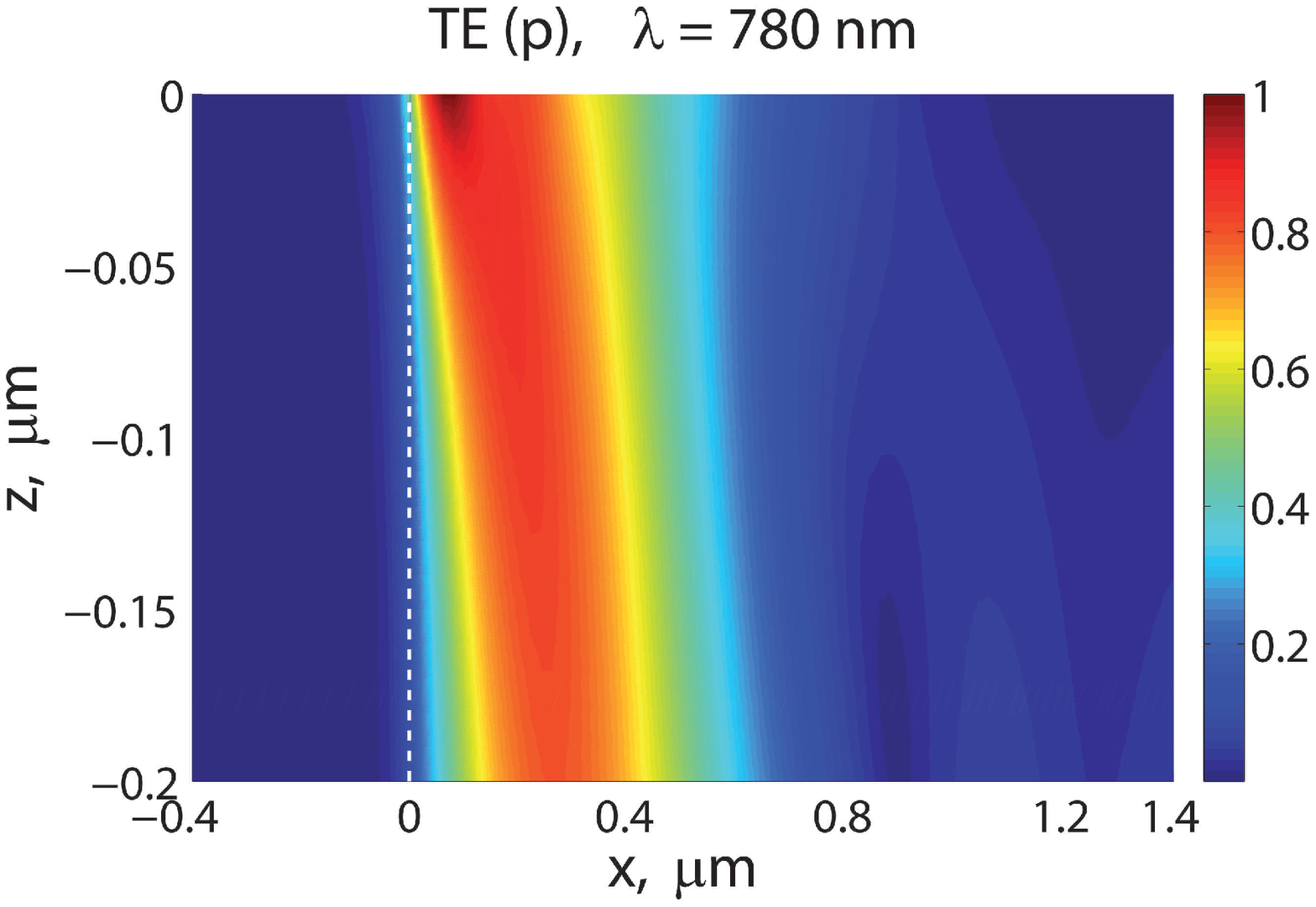} \text{(c)}
\includegraphics[scale=0.3]{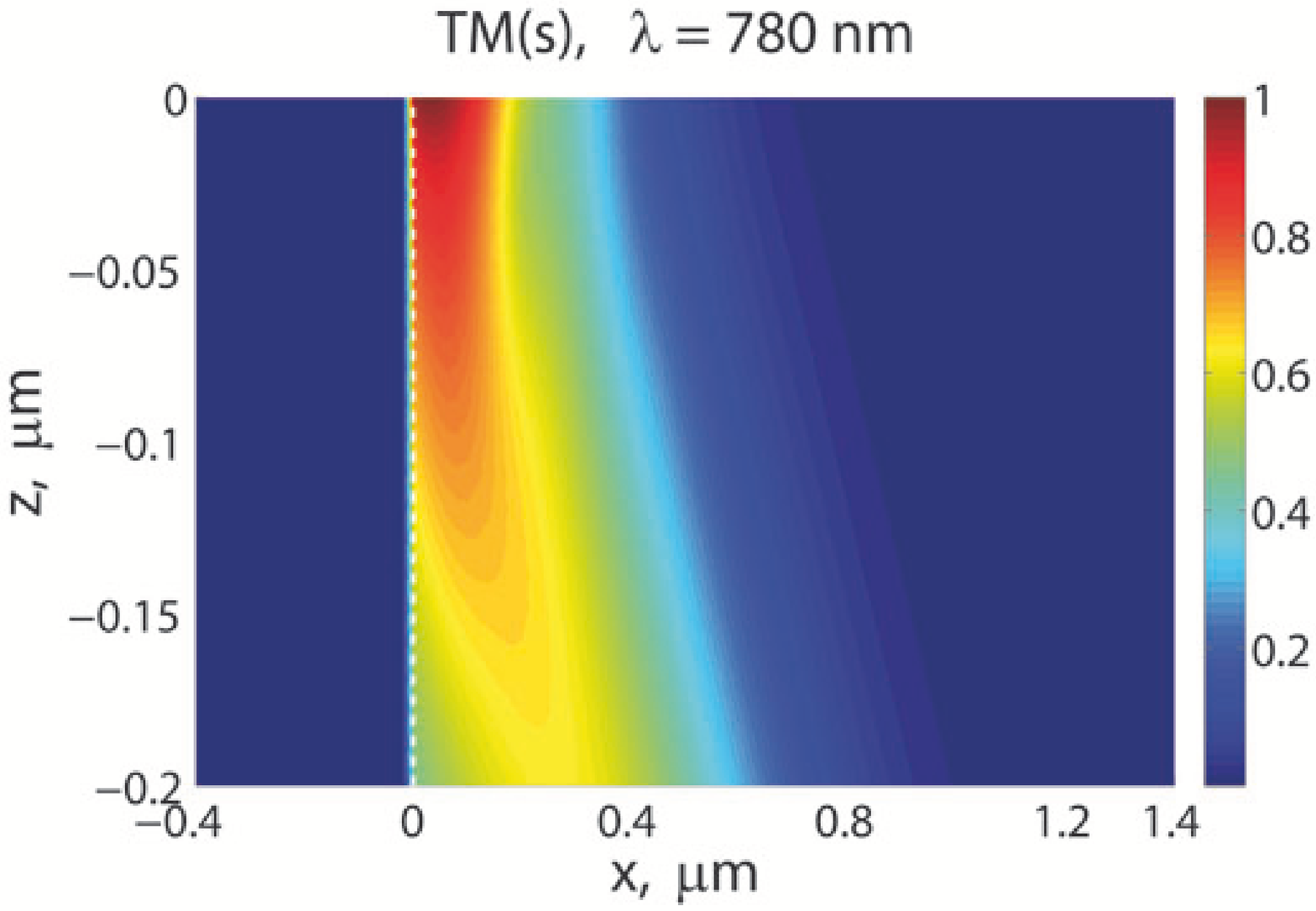} \text{(d)}
\caption{Absolute value of the electric field $\left|\mathbf{E} \right|/\left|\mathbf{E}_0 \right|$ distribution on one of the $l=2$ $\mu$m wide slit walls for TE (a,c) and TM (b,d) radiation at $\lambda = 532$ nm (a,b) and $780$ nm (c,d).  The center of the incident beam is at $x=0$. The Au knife-edge is situated at $x<0$ and the beam propagates in negative $z$-direction. The parameters of the model are $w_{0p}=0.9w_{0s}=0.55 \lambda$. }
\label{fig:fields}
\end{figure}
\begin{figure}[t!]
\centering
\includegraphics[scale=0.2]{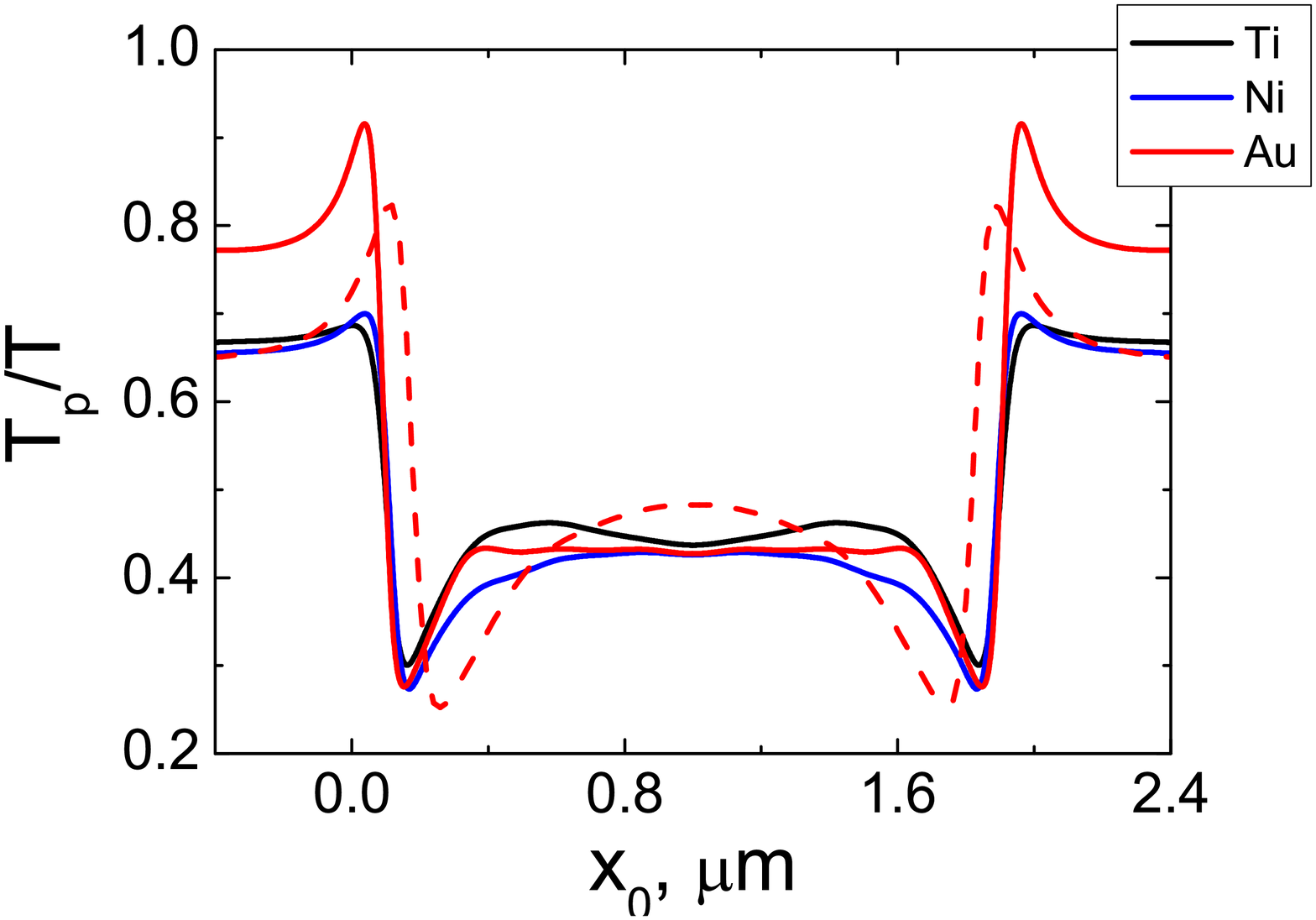} \text{(a)}
\includegraphics[scale=0.2]{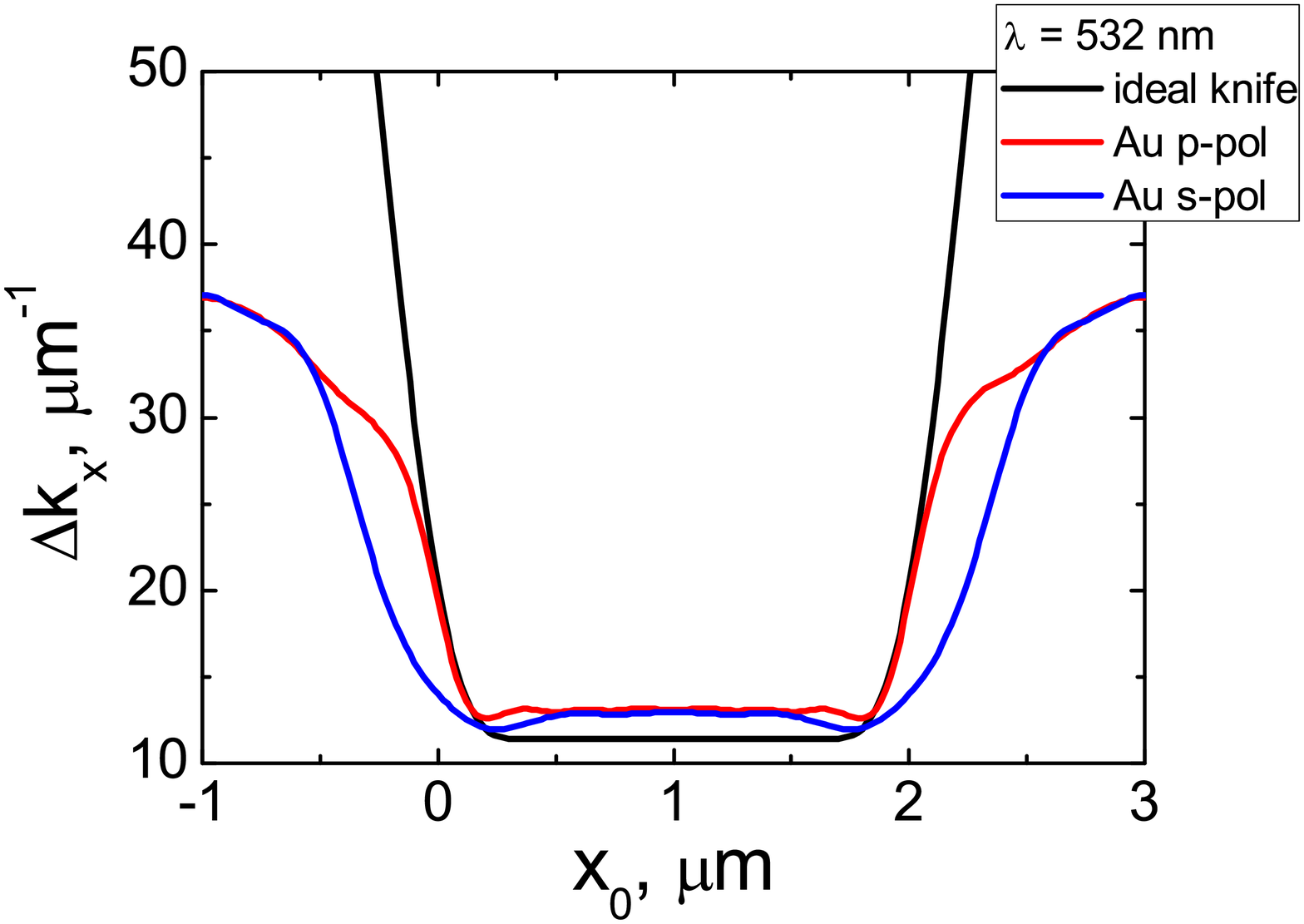} \text{(b)}
\includegraphics[scale=0.2]{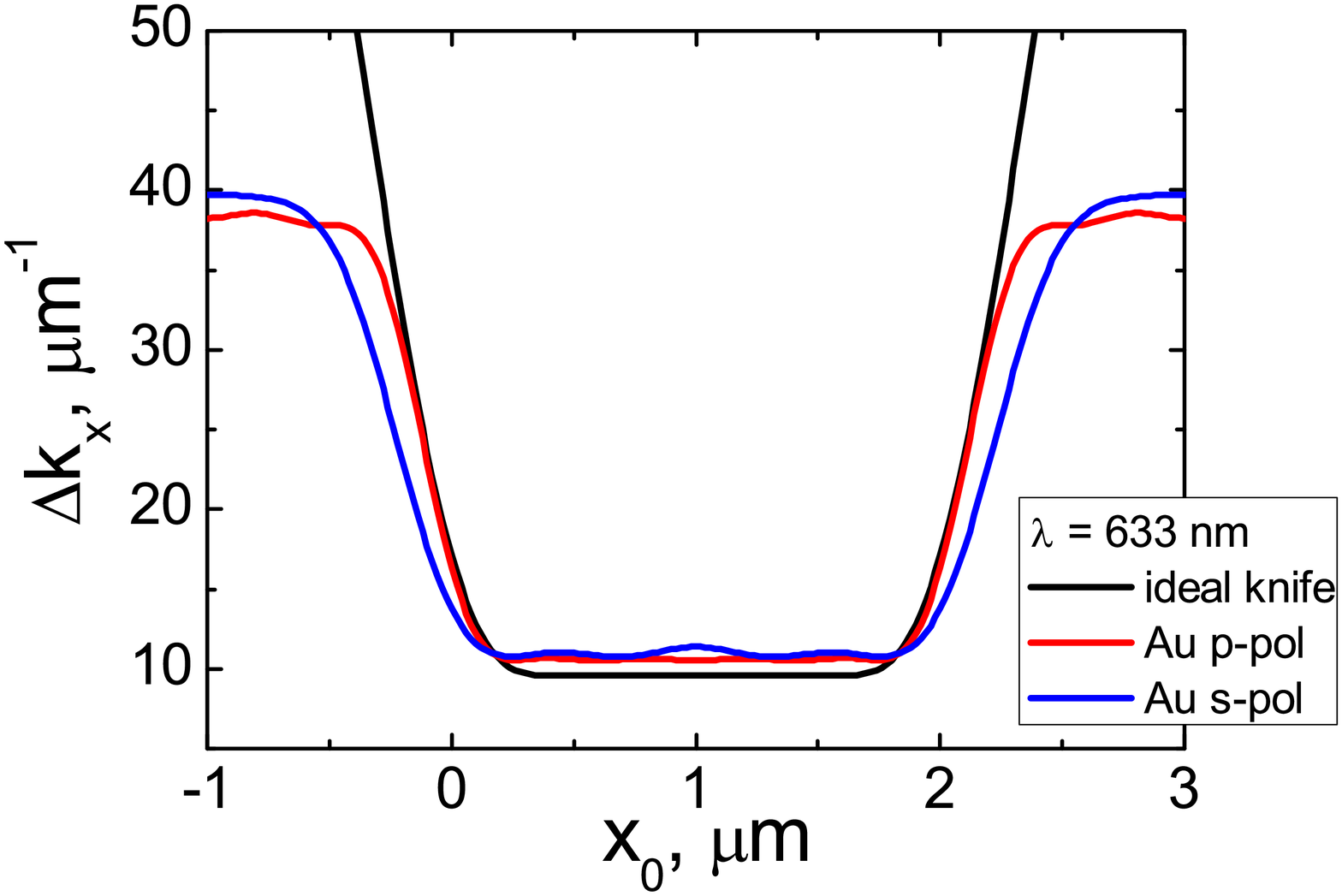} \text{(c)}
\includegraphics[scale=0.2]{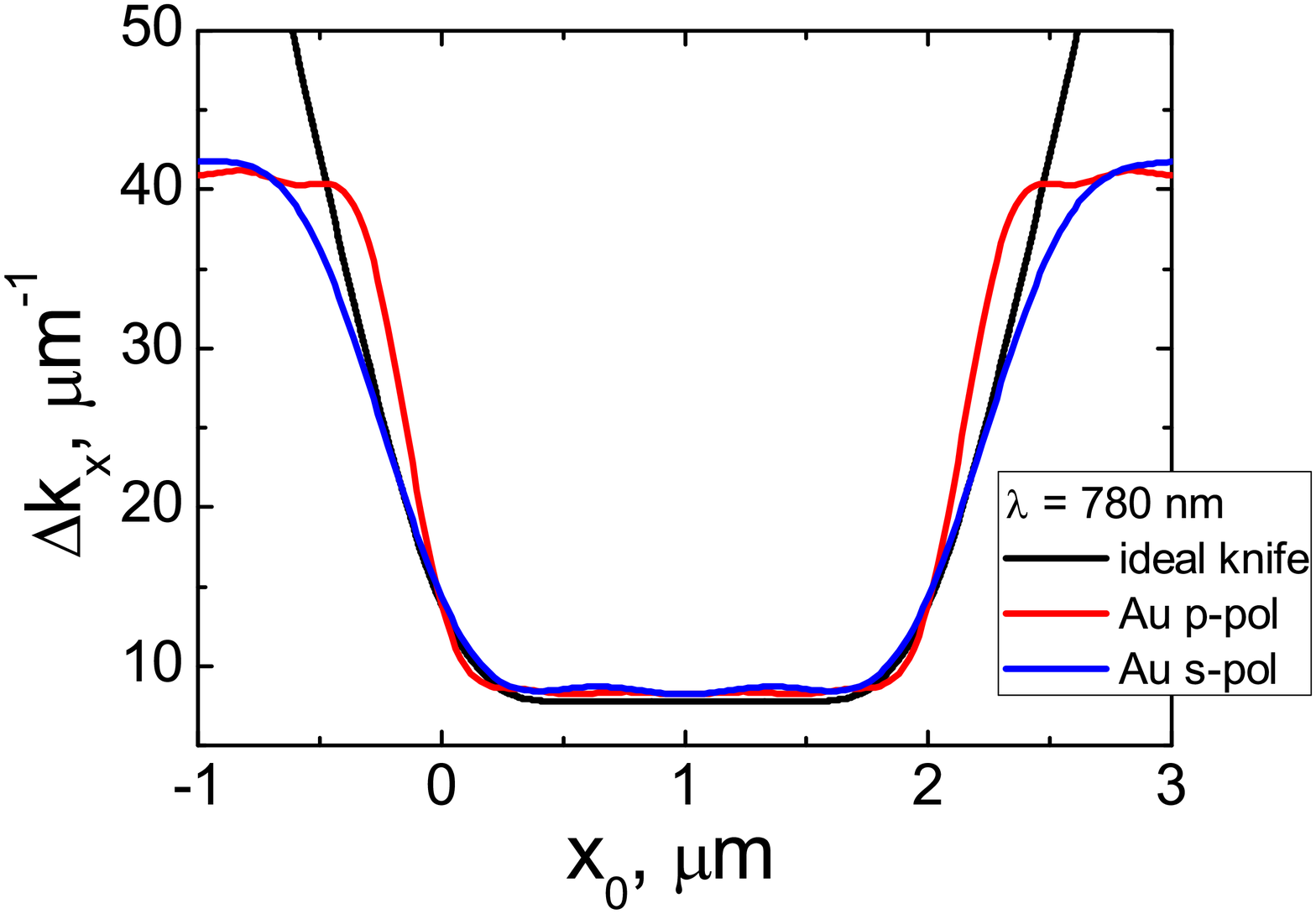} \text{(d)}
\caption{a) Fraction of energy transmitted by plasmons $T_p/T$ as a function of the beam displacement $x_0$ for various samples of the same thickness $h=200$ nm and width $l=2$ $\mu$m at $\lambda = 532$ nm (solid lines) and $\lambda = 780$ (dashed lines). b-d) Dependence of the angular spectral width $\Delta k_{x}$ of the transmitted signal's spectra $S_{trans}$ on the beam displacement $x_0$. Thickness of the Au film is $h=200$ nm, the slit width is $l=2$ $\mu$m, the wavelengths are $\lambda = 532$ nm (b), $633$ nm (c), $780$ nm (d). The spectral FWHM $\Delta k_{x}$ is determined by fitting a Gaussian function $S_{fit}\left(k_x,x_0\right)=S_0\exp\left[-2\ln2\left(k_x-k_{x0} \right)^2/\Delta k_x^2 \right]$ to the spectral distribution $S_{trans}\left(k_x,x_0\right)$. For comparison the black line represents the same dependence for an perfect knife-edge. Parameters of the theory are $w_{0p}=0.9w_{0s}=0.55 \lambda$.}
\label{fig:transmissed}
\end{figure}

\subsection{Boundary value problem at the interfaces and energy flow between regions}
In this section we discuss the boundary value problem at the entrance and exit interfaces of the slit. The eigenfunctions of the Helmholtz equation in regions I and II are those of free space, namely plane waves. On the other hand the eigenfunctions (\ref{eq:TEansatz}) of region II have discrete eigenvalues.  We note, that for the optimal convergence to be achieved, the continuity of the tangential components has to be tested differently for TE and TM polarizations \cite{ARob87}. However, for the sake of brevity, we employ here an unified and polarization-independent approach. To match the fields in the three regions the problem will be investigated in the Fourier space.

We assume, that the electric (for TE case) or magnetic (for TM case) fields in regions I and III can be expressed as
\begin{equation}
	F^{I,III} \left( x,z \right) = \frac{1}{\sqrt{2 \pi}} \int _{-\infty}^{\infty} S\left(k_x, z \right) \exp \left( \mathrm{i}k_xx \right )dk_x,
	\label{eq:radiation}
\end{equation}
here $S\left(k_x, z \right)$ is a Fourier image of the field. The slit is illuminated from the top. Therefore in the region above the first interface, the spatial spectra $S$ can be expressed as consisting of two parts. One part describes the field propagating into the slit $S_{in}$, the second part the field reflected from the slit $S_{ref}$
\begin{equation}
	S\left(k_x,z \right) = S_{in}\left(k_x \right) \exp \left[-\mathrm{i} \beta _1 \left( k_x \right) z \right]+S_{ref}\left(k_x \right) \exp \left[\mathrm{i} \beta _1 \left( k_x \right) z \right].
	\label{eq:First_interface}
\end{equation}
Here $\beta _1=\sqrt{\epsilon _{1}k^2-k_x^2}$ is the longitudinal component of the wave vector and $\mathrm{Re}\beta \geq 0$, $\mathrm{Im}\beta  \geq 0$.
We assume an incoming wave with Gaussian distribution of the $y$ component in the transverse plane. So at the entrance the field $E_y$ ($H_y$) is
\begin{equation}
	F^{G} \left( x, z = 0 \right) = \exp \left[-\left(x - x_0 \right)^2/W_0^2 \right ],
	\label{eq:Gauss}
\end{equation}
here $W_0$ is the beam waist radius and $x_0$ is the displacement from the center of the coordinates, which in our case coincides with the left wall. The FWHM $w_0$ can be found as $w_0= \sqrt{2 \ln 2} W_0$. We note, that the FWHM $w$ of the total electric field for the TM polarization is larger than $w_0$ due to the presence of the $E_z$ component. The spatial Fourier spectra of the field (\ref{eq:Gauss}) is
\begin{equation}
	S_{in}\left( k_x \right)= \frac{W_0}{\sqrt{2}}\exp\left(-\mathrm{i}k_xx_0 \right)  \exp \left(-k_x^2W_0^2/4 \right).
	\label{eq:Gaussspectra}
\end{equation}

In region III light only propagates from the slit into the substrate $S_{trans}$
\begin{equation}
	S\left(k_x,z \right) = S_{trans}\left(k_x \right) \exp \left[-\mathrm{i} \beta _3 \left( k_x \right) z \right],
	\label{eq:Second_interface}
\end{equation}
here $\beta _3=\sqrt{\epsilon _3k^2-k_x^2}$.

The electric and magnetic fields in region II consist of forward and backward propagating eigenmodes $\pi _{\nu}$. Thus, the resulting field is represented by the sum 
\begin{equation}
	F^{II} \left( x, z \right) = \sum ^{\infty}_{{\nu}=1}\left[ a_{\nu} e^{-\mathrm{i}\beta_{\nu} z}+ b_{\nu} e^{\mathrm{i}\beta_{\nu} \left(z+h \right)}\right]\pi _{\nu} \left(x \right) = \sum ^{\infty}_{{\nu}=1}F_{\nu}\left(z\right)\pi _{\nu} \left(x \right),
	\label{eq:radiation2}
\end{equation}
with the expansion coefficients $a_{\nu}$, which describe a field propagating down to the substrate, whereas the $b_{\nu}$  represent a reflected field. 

The energy conservation law can be written as 
\begin{equation}
P_{beam} = P_{trans} + P_{ref} + P_{abs},
\label{eq:Energylaw}
\end{equation}
where $P_{beam}$ is the power of the incoming beam. $P_{trans}$ is the power transmitted into the region III through the slit. $P_{ref}$ is the power reflected from the slit and $P_{abs}$ is the power losses in the slit. After some math the reflection $R$ from the slit and transmission $T$ through the slit can be expressed in terms of the spatial spectra of incoming and outgoing fields  as follows
\begin{equation}
	R = \frac{\int ^{k_{01}}_{-k_{01}} \beta _1 \left(k_x \right) \left |S_{ref}\left(k_x \right) \right| ^2 dk_x}{\int ^{k_{01}}_{-k_{01}} \beta _1 \left(k_x \right) \left |S_{in}\left(k_x \right) \right| ^2 dk_x}, \quad T =  \frac{ \mathrm{Re}(\upsilon _3) \int ^{k_{03}}_{-k_{03}} \beta _3 \left(k_x \right) \left |S_{trans}\left(k_x \right) \right| ^2 dk_x}{\mathrm{Re}(\upsilon _1) \int ^{k_{01}}_{-k_{01}} \beta _1 \left(k_x \right) \left |S_{in}\left(k_x \right) \right| ^2 dk_x},
	\label{eq:ReflTrans}
\end{equation}
where $k^2_{01} = \epsilon _1k$, $k^2_{03} = \epsilon _3k$. $\upsilon _{i}=1$ for TE field and $\upsilon _{i}=\epsilon ^{-1} _{i}$ for TM field.
The expansion coefficients $a_{\nu}$ and $b_{\nu}$ of the field inside the slit along with the spatial spectra $S_{trans}$, $S_{ref}$ are the unknowns. 

The problem of finding the unknowns is solved by implying the continuity of electric and magnetic fields at the two interfaces ($z=0$ and $z=-h$). Traditionally, the procedure of the imposition of the continuity is different for TE and TM polarizations (see Ref. \cite{JSum03,OMatM06} for examples). The main reasoning behind that is an optimal convergence. Nevertheless, for the sake of brevity, we will use the same projections for two different polarizations. This will slow the convergence, but a compact and polarization independent approach can be used. 

In addition to the continuity of $E_y$ and $H_y$, the derivative of the electric field and the derivative $\epsilon ^{-1}\partial H_y/\partial z$ of the magnetic field are continuous as well. With the help of equations  (\ref{eq:radiation}-\ref{eq:First_interface}) and (\ref{eq:radiation2}) a pair of equations is obtained at the first interface ($z=0$):

\begin{align}
	S_{ref} \left( k_x \right)& = \frac{1}{\mathrm{i}\beta _1 \left(k_x \right)}\sum _{\mu}^{\infty}\frac{\partial F_{\mu}}{\partial z}\left(0 \right)\Theta_{\mu}\left( k_x \right)+S_{in}\left(k_x \right), \nonumber \\
	2S_{\nu}&= \sum _{\mu}^{\infty}\left[\mathrm{i}\frac{\partial F_{\mu}}{\partial z}\left(0\right)J^{(1)}_{\mu \nu} + F_{\mu}\left(0\right)G_{\mu \nu}\right], 	
	\label{eq:First_system_TM}
\end{align}
where the sums run from $\mu=0$ for TM and from $\mu=1$ for TE polarization to infinity.

The matching of fields at the second interface with help of Eqs. (\ref{eq:Second_interface}) and (\ref{eq:radiation2}) gives us a similar system of equations for the unknowns $S_{trans}\left(k_x \right)$ and $a_{\mu}$, $b_{\mu}$
\begin{align}
	S_{trans} \left( k_x \right)& = \exp \left[\mathrm{i}\beta _3 \left(k_x \right)h \right]\sum _{\mu}^{\infty}F_{\mu}\left(-h \right)\Upsilon_{\mu}\left( k_x \right), \nonumber \\
		0&=\sum _{\mu}^{\infty}\left[\mathrm{i}\frac{\partial F_{\mu}}{\partial z}\left(-h\right)J^{(3)}_{\mu \nu}-  F_{\mu}\left(-h\right)G_{\mu \nu}\right], 
	\label{eq:Second_system_TM}
\end{align}

where 
\begin{align}
 \Theta_{\mu}\left( k_x \right) &= \frac{\upsilon _2}{\upsilon _1}\Phi _{\mu} \left( k_x \right)  + \Psi _{\mu}\left( k_x \right),  
 & \Upsilon_{\mu}\left( k_x \right) &= \frac{\upsilon _1}{\upsilon _3}\Theta_{\mu}\left( k_x \right) ,\nonumber \\
		J^{(1)}_{\mu \nu}&=\int _{-\infty}^{\infty}\beta ^{-1}_{1} \left( k_x \right) \Theta_{\mu}\left( k_x \right) \Pi_{\nu}\left(k_x \right) dk_x, 
		 &	J^{(3)}_{\mu \nu}&=\int _{-\infty}^{\infty}\beta ^{-1}_{3} \left( k_x \right) \Upsilon_{\mu}\left( k_x \right) \Pi_{\nu}\left(k_x \right) dk_x, \nonumber \\
		S_{\nu}&=\int _{-\infty}^{\infty} S_{in}\left( k_x \right) \Pi_{\nu}\left(k_x \right) dk_x , & G_{\mu\nu}&=\delta_{\mu\nu}\int _{-\infty}^{\infty}\Upsilon_{\nu}\left( k_x \right) \Pi_{\nu}\left(k_x \right)
	\label{eq:deffin_TM}
\end{align}
with $\Pi_{\nu}$ defined by equation (\ref{eq:TE_fourier}). The integral $S_{\nu}$ is an overlap integral of the incident beam and the eigenmode of the slit and the integral $J^{(1,3)}_{\mu \nu}$ describes  optical and geometrical properties of the slit. The overlap integral defines the behavior of the reflected and transmitted radiation.

The explicit expressions of $F_{\mu}$ and $\partial F_{\mu} / \partial z$ are
\begin{align}
	F_{\mu} \left(0\right) = a_{\mu} + b_{\mu} e^{\mathrm{i}\beta _{\mu} h}, \quad &	\frac{\partial F_{\mu}}{\partial z} \left(0\right) = \mathrm{i} \beta _{\mu} \left(-a_{\mu} + b_{\mu}e^{\mathrm{i}\beta _{\mu} h} \right),\nonumber \\
	F_{\mu} \left(-h\right) = a_{\mu}e^{\mathrm{i}\beta _{\mu} h} + b_{\mu}, \quad & 	 \frac{\partial F_{\mu}}{\partial z} \left(-h\right) = \mathrm{i} \beta _{\mu} \left(-a_{\mu}e^{\mathrm{i}\beta _{\mu} h} + b_{\mu} \right).
	\label{eq:Fn_TE}
\end{align}
We substitute those expressions into the second equations of the equations system (\ref{eq:First_system_TM}-\ref{eq:Second_system_TM}) and get
\begin{align}
	2S_{\nu} = \sum _{\mu} \left(G_{\mu\nu} + \beta_{\mu}J^{(1)}_{\mu\nu} \right)  a_{\mu} +e^{\mathrm{i}\beta _{\mu} h}\left( G_{\mu\nu} - \beta_{\mu}J^{(1)}_{\mu\nu}\right)b_{\mu},\nonumber \\
	0 = \sum _{\mu}e^{\mathrm{i}\beta _{\mu} h} \left(G_{\mu\nu} - \beta_{\mu}J^{(3)}_{\mu\nu} \right)a_{\mu}  +\left( G_{\mu\nu} + \beta_{\mu}J^{(3)}_{\mu\nu}\right)b_{\mu}.
	\label{eq:final_TE_system}
\end{align}

Now, we rewrite Eq. (\ref{eq:final_TE_system}) in the following matrix form 
\begin{equation}
	2\mathbf{S}=\mathbf{N}_a^{(1)}\mathbf{a}+\mathbf{N}_b^{(1)}\mathbf{b}, \quad
	0=\mathbf{N}_a^{(2)}\mathbf{a}+\mathbf{N}_b^{(2)}\mathbf{b},
	\label{eq:matrixsystem_TM}
\end{equation}
with the matrices $\mathbf{N}$ being defined as
\begin{align}
	\mathbf{N}_a^{(1)}&=\mathrm{diag}\left(\beta _{\mu} \right)\mathbf{J}^{1}+\mathbf{G}, & 
	\mathbf{N}_b^{(1)}&=\mathrm{diag}\left(e^{\mathrm{i}\beta _{\mu} h} \right)\left[\mathbf{G}-\mathrm{diag}\left(\beta _{\mu} \right)\mathbf{J}^{1}\right], \nonumber \\
	\mathbf{N}_a^{(2)}&=\mathrm{diag}\left(e^{\mathrm{i}\beta _{\mu} h} \right)\left[\mathrm{diag}\left(\beta _{\mu} \right)\mathbf{J}^{3}-\mathbf{G}\right], &
	\mathbf{N}_b^{(2)}&=-\mathrm{diag}\left(\beta _{\mu} \right)\mathbf{J}^{3}-\mathbf{G}. 
	\label{eq:matrices_TM}
\end{align}
They solve the resulting matrix equations for unknown vectors $\mathbf{a}$ and $\mathbf{b}$. 

Now, from Eqs. (\ref{eq:First_system_TM}) and (\ref{eq:Second_system_TM}), we obtain the unknown Fourier images $S_{trans}$ and $S_{ref}$. In particular, the transmitted power $T$, see (\ref{eq:ReflTrans}), can be expressed as

\begin{equation}
T =  \frac{ \mathrm{Re}\upsilon _3 }{\mathrm{Re}\upsilon _1 P_{beam}}\mathrm{Re} \sum _{\mu}^{\infty}\sum _{\nu}^{\infty} \mathrm{i}\frac{\partial F_{\mu}}{\partial z} \left(-h\right)  F^{*}_{\mu} \left(-h\right)H_{\mu \nu}^{(2)},
\label{eq:Transm}
\end{equation}
where 
\begin{equation}
H_{\mu \nu}^{(2)} = \int_{-k_{03}}^{k_{03}} \Upsilon_{\mu}\left( k_x \right)\Pi^{*}_{\nu}\left(k_x \right)  \mathrm{d}k_x.
\label{eq:H}
\end{equation}

The physical meaning of the derived formulation can be explained as follows. The propagation constants $\beta _{\mu}$ are complex, so each eigenmode is attenuated while it propagates from the entrance to the exit. The incoming field excites eigenmodes in the slit. The strength of the excitation is defined by the overlap of the eigenmode and the incoming field given by a certain distribution of transverse wave vectors $k_x$. This means, that the slit acts as a complex spatial filter, acting differently on the spatial frequencies $k_x$ according to the damping of the respective mode excited in the slit. The higher the value of the initial $k_x$ the stronger is the interaction of the light with the slit walls. Those modes are damped even stronger for deeper slits. If a certain thickness $h$ is exceeded only the first mode will reach the exit of Region II, effectively leaving mainly two spatial frequencies $k_x=\pm\mathrm{i}\gamma_1$, $k_x=\pm\mathrm{i}\gamma_2$ in the transmitted signal (see Eq. (\ref{eq:TE_fourier})). This holds true in general for both polarizations with one notable difference. For s-polarization besides the propagating modes there are two plasmonic modes, which are exponentially localized at the slit walls. For narrow slits the plasmons are damped less than all the other modes. Hence, the electric field distribution at the photodiode may develop two distinct spatial peaks - each at it's wall of the slit.

\subsection{Numerical simulations and discussion}
In the following section we present and discuss the results of our numerical simulations based on our theoretical model. For a given value of the beam displacement $x_0$ the procedure of finding the unknowns $\mathbf{a}$ and $\mathbf{b}$ (see Eq. (\ref{eq:matrixsystem_TM})) was as follows. As a first step, roots of Eqs. (\ref{eq:TEequa}) were calculated by built-in iterative MatLab procedures. The values of the propagation constant $\beta _{\nu}$ for a perfectly conducting wall were used as an initial guess. For plasmonic modes, an initial guess was the propagation constant $\beta_p=\frac{\omega}{c_0}\left[\epsilon_2/\left(1+\epsilon_2\right)\right]^{1/2}$ of a single plasmonic mode propagating at the interface between the knife-edge walls and the slit. The matrix dimensions were chosen in such a way that the energy conservation criterion (\ref{eq:Energylaw}) was satisfied with a relative accuracy of $10^{-8}$. In general, the typical length of the unknown vectors was about $40$-$50$ elements due to the beamwidth of the incident field close to the wavelength. All dielectric constants were taken from \cite{MJWeb}. We carefully choose free parameters of the theory ($w_{0p}=0.9w_{0s}=0.55 \lambda$) to achieve the best possible agreement with our experimental data. The FWHM of the squared electric fields in the model are $w_{p}=0.8w_{s}$ and slightly differ from the expected in our setup \cite{BRic59}. 

As a numerical result the electric field distribution at a knife-edge maid from gold is presented in Fig. \ref{fig:fields} for two different wavelengths and two polarizations (s and p). We remind here, that for small values of $z$ the electric field distribution is approximate, see Section 3.2. The beam is impinging from the top and the thickness of the knife-edge is $h=200$ nm. The observed patterns are similar to those observed during the diffraction of the beam from a wedge, see for example \cite{YZUmu07}. For both states of polarization beam diffraction is observed (see Fig. \ref{fig:fields}). In the case of an s-polarized incoming beam, the excitation of plasmonic modes at the edge is expected and can be seen in Fig. \ref{fig:fields} (b,d). In all graphs we see a shift of the electric field maximum introduced by reflection from the side wall of the knife surface, as already suggested in the discussion in Section 3.1. The beam reaches the photodiode at different positions. In all cases some part of the beam penetrates the knife-edge and reaches the photodiode as well. As a main difference, the part of the field attributed to the plasmonic modes is largest at $\lambda=780$ nm for s-polarization, whereas at $\lambda=532$ nm the plasmonic modes are attenuated comparatively fast due to increased absorption. For that reason localization of an s-polarized field is less strong in the latter case.

In addition to the Au sample we consider Ni and Ti samples of the same thickness ($h=200$ nm) and illustrate the importance of plasmonic modes in the transmission of the TM fields (s-polarized) at next. The amount of power transmitted by the plasmonic modes as a function of beam displacement is plotted in Fig. \ref{fig:transmissed} (a). We see that due to their spatial extent into the metal knife-edge plasmonic modes have a noticeable influence on the transmission, even, if most of the beam is blocked. So, for a gold film at $\lambda = 532$ nm they account for up to $90\%$ of the transmitted power for a beam impinging on the slit walls. For Ni and Ti samples this value decreases to nearly $70\%$, while at $\lambda = 780$ nm the Au-knife-edge maintains it's plasmonic properties. The last point we have to note is that still a non-zero amount of energy is transmitted by plasmons even for the beam centered in the slit. This happens due to the excitation of the coupled plasmons pair on both walls of the slit. 

As a further result of our theoretical calculations we see that a knife edge acts as a filter of the incoming beams spatial spectrum. To have a reference we start with a pair of perfect knife-edges, see Eq. (\ref{eq:knife}), for which the spatial spectrum of the transmitted signal is $S_{trans}\left(k_x,x_0\right)=\int_{0}^{l}F^{G}\left(x\right)\exp(-\mathrm{i}k_xx) \mathrm{d}x$. For convenience, we determine the spectral FWHM $\Delta k_x$ by fitting a Gaussian function $S_{fit}\left(k_x,x_0\right)=S_0\exp\left[-2\ln2\left(k_x-k_{x0} \right)^2/\Delta k_x^2 \right]$ to the resulting spectral distribution $S_{trans}\left(k_x,x_0\right)$. The spectral profile remains symmetric for an perfect knife-edge and the dependence of the angular width $\Delta k_x$ is presented in Fig. \ref{fig:transmissed} (b-d). Our intention here is to compare the spectral width of the signal chopped by the real knife-edge to the spectral FWHM of the same signal chopped by the perfect knife-edge. During beam profiling with a perfect knife-edge the spatial spectrum broadens, when the beam is blocked by the knife-edge. However, the  spectral width of the signal blocked by the other knife-edges is slightly larger compared to the perfect case, when the beam is in the middle of the slit. Nevertheless, as the beam approaches the knife-edge, the spectral width becomes much smaller than expected for a perfect knife-edge. This is a direct indication of the filtering of spatial frequencies. So, dependent on the polarization the reconstruction may result in a beam profile wider than expected, if other effects can be neglected. 

\section{Conclusions}
In conclusion, the knife-edge method at nanoscale was investigated in detail. We experimentally as well as theoretically investigated the knife-edge technique for reconstructing the intensity profile of highly focused linearly polarized beams. Different thicknesses and edge materials were studied. It was demonstrated that a variety of polarization sensitive physical effects can influence the successful beam reconstruction: The excitation of plasmonic modes, the polarization dependent reflection from the wall of the knife-edge and a non-zero power-flow through the edge material. Thus, the reconstructed beam profile strongly depends on the used material, the wavelength and the thickness of the knife-edge. The peak position of the conventionally reconstructed beam as well as its width is affected by the interaction of the knife-edge with the measured beam. For the fabrication of knife-edges for highly accurate beam reconstruction a suppression of plasmonic modes is crucial along with minimization of a polarization dependent reflection from the inner side of the knife-edge. For pure materials the suggested workaround can be the use of very thin metal films. Another possibility to minimize the influence of the plasmonic modes and the polarization dependent reflection is the use of films with rough internal structure, where the reflection is diffuse and plasmons are suppressed (compare  \cite{RDorn03a,RDorn03b}).

\section{Acknowledgments}
The support by the DFG funded Erlangen Cluster of Excellence "Engineering of Advanced Materials" (EAM) is gratefully acknowledged. We thank Stefan Malzer, Isabel G\"a\ss ner, Olga Rusina, Irina Harder and Daniel {Plo\ss} for their valuable support in preparing the samples. S. Orlov acknowledges the support by the Humboldt Foundation. 
\end{document}